\documentclass[aps,pra,reprint,amsmath,amssymb,nofootinbib,longbibliography,floatfix,superscriptaddress]{revtex4-2}
\usepackage{mathtools,bm}

\usepackage{amsthm}
\usepackage{graphicx,booktabs,xcolor}
\usepackage{microtype}
\usepackage[colorlinks=true,linkcolor=blue!45!black,citecolor=blue!45!black,urlcolor=blue!45!black]{hyperref}
\usepackage{cleveref}
\crefname{equation}{Eq.}{Eqs.}
\Crefname{equation}{Eq.}{Eqs.}
\allowdisplaybreaks[2]
\newtheorem{theorem}{Theorem}
\newtheorem{proposition}{Proposition}
\newtheorem{lemma}{Lemma}
\newtheorem{corollary}{Corollary}
\theoremstyle{remark}

\DeclareMathOperator{\Tr}{Tr}
\DeclareMathOperator{\supp}{supp}

\DeclareMathOperator{\Bin}{Bin}
\DeclareMathOperator{\SEP}{SEP}

\DeclareMathOperator{\diag}{diag}
\DeclareMathOperator{\Li}{Li}
\newcommand{\Cspec}{\mathcal C_{\mathcal B,\eta}}
\newcommand{\Bspec}{B_{\mathcal B,\eta}}
\newcommand{\Bfade}{B_\mu^{\mathrm{fad}}}
\newcommand{\id}{\mathrm{id}}
\newcommand{\E}{\mathbb E}
\newcommand{\Prb}{\mathbb P}

\newcommand{\Leta}{\mathcal L_{\boldsymbol\eta}}
\newcommand{\Feta}{F_{\boldsymbol\eta}}
\newcommand{\Beta}{B_{\boldsymbol\eta}}
\newcommand{\ER}{E_R}
\newcommand{\ERi}{E_R^\infty}

\newcommand{\ket}[1]{\lvert #1\rangle}
\newcommand{\bra}[1]{\langle #1\rvert}
\newcommand{\norm}[1]{\left\lVert#1\right\rVert}
\newcommand{\dd}{\,\mathrm d}
\definecolor{capblue}{RGB}{20,80,140}
\definecolor{caporange}{RGB}{182,80,30}
\definecolor{capgreen}{RGB}{35,130,105}

\newcommand{\F}{\mathcal F_\mu}
\newcommand{\Fhat}{\widehat{\mathcal F}_\mu}
\newcommand{\Crec}{C_{\mathrm{rec}}}
\newcommand{\Cpre}{C_{\mathrm{pre}}}
\newcommand{\Afun}{\mathsf A_\mu}
\newcommand{\Bfun}{\mathsf B_\mu}
\newcommand{\proj}[1]{\ket{#1}\!\bra{#1}}

\begin{document}
\title{Finite-energy two-way assisted capacities: \\ Exact formulas for multimode loss and fading}
\author{Stefano Pirandola}
\affiliation{Department of Computer Science, University of York, York YO10 5GH, United Kingdom}
\author{Gaetana Spedalieri}
\affiliation{nodeQ, 71-75 Shelton
Street, Covent Garden, London WC2H 9JQ, United Kingdom}

\begin{abstract}
We study quantum communication, entanglement distribution, and
secret-key generation through multimode bosonic loss and fading channels,
assisted by local quantum operations and two-way classical
communication. Using sector teleportation, we determine the exact
two-way quantum and secret-key capacities of multimode pure-loss
channels under a mean photon-number constraint, allowing arbitrary
adaptive quantum protocols. We extend these results to continuous
frequency-dependent loss under a power constraint and identify the
optimal allocation of photons among optical modes. For constant loss
over an unlimited bandwidth, we obtain a closed-form capacity
proportional to the square root of power. For randomly varying loss,
we determine exact finite-energy capacities when the channel
realization is publicly revealed, distinguishing knowledge before
and after transmission. When fading remains unobserved, we derive
capacity bounds and explicit formulas for an on/off channel.
We also establish strong-converse bounds, with exact thresholds
for the deterministic loss models considered here, observed fading,
and unobserved fading with finitely many possible transmissivities.
Under mean energy constraints, these thresholds generally exceed
the corresponding finite-energy capacities. These results guide energy
allocation and provide benchmarks for energy-constrained optical quantum
communication with multimode signals and variable loss.
\end{abstract}
\maketitle

\section{Introduction}\label{sec:intro}

Optical communication uses spatial, polarization, temporal, and
frequency modes with generally unequal losses. A finite energy budget
must be allocated among these modes, while a capacity converse must
exclude improvements from correlated inputs, collective decoding,
and adaptive protocols. In this context, bosonic Gaussian channels describe attenuation
and mode mixing \cite{HolevoWerner,Weedbrook,Serafini2017} and passive optical
systems admit a description in terms of independent attenuators
\cite{Lupo}.

For a pure-loss channel with transmissivity $\eta$, the unconstrained
two-way capacity is the PLOB rate $-\log_2(1-\eta)$ \cite{PLOB}.
At finite mean input photon number $N$, reverse coherent information
(RCI) yields the achievable rate $g(N)-g((1-\eta)N)$
\cite{Reverse,DirectReverse}. Recently, by using sector-teleportation simulation,
Ref.~\cite{Pirandola2026} established the optimality of this rate
for arbitrary adaptive protocols under an unconditional mean
transmitted-photon-number constraint. In this work, we extend the results of Ref.~\cite{Pirandola2026} to multimode signals and, later, to fading channels.

Using the finite-energy converse of Ref.~\cite{Pirandola2026}, we
determine the exact multimode capacities for arbitrary adaptive
protocols. The capacity is obtained by optimizing the photon
allocation across modes. At a positive budget, the optimal allocation
is unique and assigns photons to every transmitting mode. For passive
optical systems under a total photon-number constraint, the capacity
depends only on the squared singular values of the transmission
matrix. We also treat weighted energy constraints when the energy
cost is diagonal in the loss-mode basis.

We then consider a continuous frequency-dependent loss profile under
a power constraint. Building on the broadband frameworks of
Refs.~\cite{CavesDrummond1994,Broadband}, we derive an exact spectral allocation formula
and justify the continuum limit, including the low- and high-frequency
tails. For constant loss over an unlimited band, we evaluate the
capacity analytically, obtaining a square-root power law.

We further study fading channels, whose transmissivity fluctuates
because of beam wandering, turbulence, pointing errors, or obstruction
\cite{Pirandola2021,Vasylyev2012,Vasylyev2016,Papanastasiou2018}.
Their capacities depend on whether the instantaneous transmissivity
is revealed or remains unobserved, with practical access to this
information governed by channel tracking and receiver temporal
resolution. Ensemble averages of the PLOB rate underlie free-space
and satellite communication limits
\cite{Pirandola2021,PirandolaSatellite2021}. Pilot-assisted tracking
and receiver trust assumptions also enter practical security analyses
\cite{PirandolaTrust2021}, while complementary bounds
connect the tolerable quantum bit error rate to the maximum distance
of free-space quantum key distribution \cite{PirandolaQBER2026}.
At finite energy, we determine exact capacities when the fading
realization is revealed before or after transmission. For unobserved
fading, we derive finite-energy upper and lower bounds, with
scalar formulas for binary on/off fading.

Finally, we study the rates above which communication failure becomes
asymptotically certain under mean energy or power constraints.
We determine these strong-converse thresholds for the multimode and broadband models, as well as for observed fading and unobserved fading
with finitely many possible transmissivities. For general unobserved
fading with a finite average PLOB rate, we establish an upper bound
on the threshold. Under these mean constraints, the strong-converse
thresholds generally exceed the corresponding finite-energy capacities. 

We first treat multimode loss, including continuous-frequency profiles,
and then memoryless fading.
The proofs appear in Appendices~\ref{app:structure}--\ref{app:endpoints}
and~\ref{fd:app:map}--\ref{fd:app:strong}, respectively.

\vspace{-0cm}
\section{Channel model and operational constraints}\label{sec:model}
Fix $m<\infty$ and initially assume $0<\eta_j<1$ for $j=1,\ldots,m$.
One use of the channel is
\begin{equation}
\Leta=\bigotimes_{j=1}^{m}\mathcal L_{\eta_j},~~
 q_j=1-\eta_j,~~
 b_j=\sqrt{\eta_j}\,a_j+\sqrt{q_j}\,e_j,
 \label{eq:product}
\end{equation}
where the environment modes $e_j$ are in the vacuum. Different uses have
fresh vacuum environments. (The endpoints $\eta_j=0,1$ are specifically treated in Appendix~\ref{app:endpoints}.)

The admissible resources are local state preparation, quantum storage,
and an authenticated public classical link in both directions. Publicly
known randomness may coordinate the parties' initial preparations, but
no entangled state or secret key is supplied in advance. A finite
protocol consists of a finite sequence of local operations whose
classical records take values in finite or countable sets. The
adversary has access to every environmental output of the loss channels
and to every message exchanged over the public link. Local registers,
including any private-state shields, are modeled on arbitrary
separable Hilbert spaces and need not obey an energy bound. The
constraint is imposed solely on the optical systems sent through the
channel.

For a protocol using \eqref{eq:product} $n$ times, let
$\rho^{(i)}_{A_{i1}\cdots A_{im}}$ be its unconditional input at round $i$.
Set
\begin{equation}
 x_{ij}=\Tr(\hat n_{A_{ij}}\rho^{(i)}_{A_{ij}}),
 ~~
 \frac1n\sum_{i=1}^{n}\sum_{j=1}^{m}x_{ij}\le N .
 \label{eq:energy}
\end{equation}
The scalar $N$ is the total photon budget per \emph{multimode use}.
Every branch, random choice, and unsuccessful outcome contributes to the
mean, and all channel slots count toward the rate. No branchwise peak
constraint is imposed. In particular, the budget in \eqref{eq:energy}
is not a budget of $N$ photons for each of the $m$ modes.

The capacities $Q_2,D_2,P_2,K_2$ refer respectively to quantum
transmission, entanglement distribution, private classical transmission,
and secret-key agreement. They are measured in qubits, ebits, or secret
bits per multimode use. The quantum error is entanglement infidelity. For secret-key generation, let $\kappa_{K_AK_BEZ}$ denote the
protocol's unconditional final state, where $K_A$ and $K_B$ are
Alice's and Bob's classical key registers, $E$ contains Eve's quantum
systems, and $Z$ records the entire public transcript. Any remaining
local memories are traced out. For key registers with alphabet
$\{0,\ldots,d-1\}$, we define the operational error as
\begin{equation}
 \begin{gathered}
 e=\frac12\norm{\kappa_{K_AK_BEZ}
 -\tau_d^{\mathrm{key}}\otimes\kappa_{EZ}}_1,\\
 \tau_d^{\mathrm{key}}
 =\frac1d\sum_{a=0}^{d-1}\ket{aa}\bra{aa},
 \end{gathered}
 \label{eq:keyerror}
\end{equation}
where
$\kappa_{EZ}=\Tr_{K_AK_B}\kappa_{K_AK_BEZ}$
is the actual marginal state of Eve's systems and the public record.
In the ideal comparison state, Alice and Bob hold identical,
uniformly distributed keys that are independent of $EZ$, whose
marginal is unchanged. Thus $e$ measures the joint deviation from
perfect key agreement and secrecy \cite{PirandolaQKD2020,PortmannRenner}. Rates are $\log_2d/n$, and capacities require error tending
to zero. Teleportation and the one-time pad give $Q_2=D_2$ and $P_2=K_2$
without changing transmitted energies.

Let us define the following functions
\begin{align}
 g(x)&=(x+1)\log_2(x+1)-x\log_2x,\label{fd:eq:g}\\
 c_\eta(x)&=g(x)-g((1-\eta)x),\label{fd:eq:c}\\
 \Feta(N)&=\max_{\substack{N_j\ge0\\\sum_{j=1}^{m}N_j\le N}}
                   \sum_{j=1}^{m}c_{\eta_j}(N_j),\label{eq:F}\\
 \Beta&=\sum_{j=1}^{m}-\log_2q_j,\label{eq:B}
\end{align}
where $\Beta$ is the unconstrained value~\cite{PLOB}. Note that the feasible set in \eqref{eq:F} is compact so the maximum exists.

\section{Exact multimode capacities}\label{sec:capacity}

\begin{theorem}[Adaptive capacities]\label{thm:capacity}
Under the unconditional constraint \eqref{eq:energy},
\begin{equation}
 \boxed{
 \begin{aligned}
 Q_2(\Leta,N)&=D_2(\Leta,N)\\
 &=P_2(\Leta,N)=K_2(\Leta,N)\\
 &=\Feta(N).
 \end{aligned}}
 \label{eq:capacity}
\end{equation}
The same identities hold under the stronger per-use unconditional
constraint $\sum_jx_{ij}\le N$ for every $i$.
\end{theorem}

The proof is given in Appendices~\ref{app:adaptive} and~\ref{app:achieve}. Its essential
steps can be stated directly. Let $\zeta_n$ denote the final retained
state of the parties, with all necessary local shields included, and
let $r$ be a number of independent replicas of this fixed protocol.
Applying the single-mode block estimate to each pair $(i,j)$ across the
replicas yields
\begin{align}
 \ER(\zeta_n^{\otimes r})
 \le{}&r\sum_{i=1}^{n}\sum_{j=1}^{m}c_{\eta_j}(x_{ij})\nonumber\\
 &+\sum_{i=1}^{n}\sum_{j=1}^{m}
       \big[g(r\eta_jx_{ij})+g(rq_jx_{ij})\big].
 \label{eq:replicamain}
\end{align}
Here $\ER$ is the relative entropy of entanglement across the complete
Alice--Bob cut \cite{VedralPlenio}. The additional terms are sublinear
in $r$ for every fixed finite input energy.

Dividing by $r$ and taking $r\to\infty$ at fixed protocol gives
\begin{equation}
 \ERi(\zeta_n)\le\sum_{i,j}c_{\eta_j}(x_{ij}).
 \label{eq:adaptivegain}
\end{equation}
No independence between physical channel inputs is required. Define
$\bar N_j=n^{-1}\sum_i x_{ij}$. Concavity in the energy then gives
\begin{equation}
 \ERi(\zeta_n)
 \le n\sum_jc_{\eta_j}(\bar N_j)
 \le n\Feta(N).
 \label{eq:master}
\end{equation}
Quantum-target and privacy tests turn this estimate into finite-error
weak converses. In particular, a quantum output of infidelity
$\varepsilon$ and a classical key of error $e$ satisfy, respectively,
\begin{align}
 (1-\varepsilon)\log_2d
 &\le n\Feta(N)+h_2(\varepsilon),\label{eq:quantweak}\\
 (1-e)^2\log_2d
 &\le n\Feta(N)+h_2(2e-e^2),\label{eq:keyweak}
\end{align}
where $h_2$ is binary entropy.

For achievability, let $N_j$ be any feasible allocation. Independent
two-mode squeezed-vacuum pairs with transmitted means $N_j$ give the
reverse coherent information
\begin{equation}
 I_{\mathrm{rev}}
 =\sum_j\big[g(N_j)-g(q_jN_j)\big].
 \label{eq:reverse}
\end{equation}
Reverse hashing attains this rate
\cite{Reverse,DirectReverse,BennettDistillation1996,DevetakWinter}.
A finite-dimensional sector-hashing construction in
Appendix~\ref{app:achieve} supplies the same limit with exact photon-number
control. It also proves the per-use statement and provides the
finite-dimensional coding ingredient for the continuum extension. Thus optimized independent inputs in the loss eigenmodes attain the
capacity of the full adaptive protocol class. Correlations in a physical
mode basis may still arise from the passive encoding transformation of
\cref{sec:passive}.

\section{Optimal photon allocation}\label{sec:allocation}

The allocation problem in \eqref{eq:F} is a strictly concave maximization.
For $0<\eta<1$, $q=1-\eta$, and $x>0$,
\begin{align}
 c_\eta'(x)
 &=\log_2(1+x^{-1})-q\log_2(1+(qx)^{-1}),\label{eq:slope}\\
 c_\eta''(x)
 &=-\frac{\eta}{\ln2\,x(x+1)(1+qx)}<0.\label{eq:curvature}
\end{align}
The derivative decreases from $+\infty$ to zero; consequently $c_\eta$
is strictly increasing.

\begin{proposition}[Unique allocation]\label{prop:allocation}
For $N>0$ there is a unique maximizing vector
$\boldsymbol N^*(N)$ in \eqref{eq:F}. It has $N_j^*>0$ for every
$j$ and $\sum_jN_j^*=N$. There is a unique $\lambda>0$ such that
\begin{equation}
 c_{\eta_j}'(N_j^*)=\lambda,~~ j=1,\ldots,m.
 \label{eq:kkt}
\end{equation}
If $\eta_j>\eta_k$, then $N_j^*>N_k^*$. The value function
$\Feta$ is increasing and concave, and
$\Feta'(N)=\lambda$ for $N>0$.
\end{proposition}

A proof, including the ordering of the optimal means, is given in
Appendix~\ref{app:allocation}. Numerically, invert each strictly decreasing
function $c_{\eta_j}'$ and adjust $\lambda$ until the means sum to $N$.
Using $\log N_j$ as the root variable resolves very small allocations
on weakly transmitting modes.

\begin{corollary}[Identical losses]\label{cor:identical}
If $\eta_j=\eta$ for all $j$, the optimal allocation is $N_j^*=N/m$ and
\begin{equation}
 \boxed{\Feta(N)=m\left[g(N/m)-g((1-\eta)N/m)\right].}
 \label{eq:identical}
\end{equation}
For a total budget $m\bar N$, the rate per multimode use is
$m\,c_\eta(\bar N)$ and the rate per optical mode is $c_\eta(\bar N)$.
\end{corollary}

Low- and high-energy expansions are given in Appendix~\ref{app:allocation}.
Figure~\ref{fig:allocation} illustrates both the capacity gain from
optimizing the allocation and the resulting modal photon fractions.

\begin{figure*}[t]
 \centering
 \includegraphics[width=\textwidth]{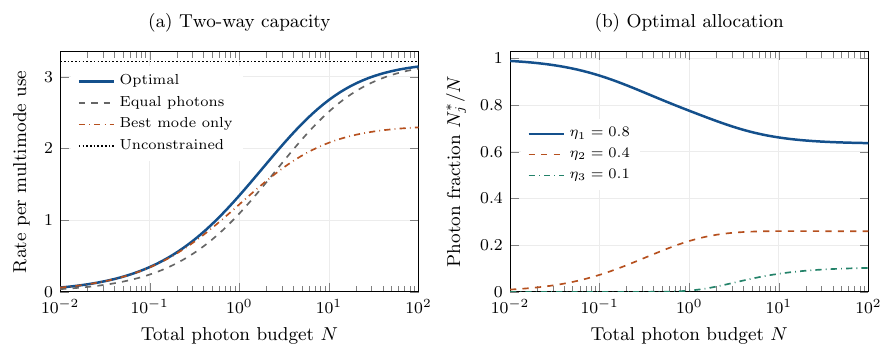}
 \caption{Three-mode example with
 $\boldsymbol\eta=(0.8,0.4,0.1)$. (a) The exact capacity from
 \eqref{eq:F} compared with two feasible allocations. The dotted line is
 $\Beta=-\log_2(0.2\cdot0.6\cdot0.9)\simeq3.21090$.
 (b) The solution of \eqref{eq:kkt}. All modal means are positive at
 finite $N>0$, although the weakest-mode fraction is very small at low
 energy. The curves approach the high-energy fractions derived in
 Appendix~\ref{app:allocation}.
 Each point is obtained by scalar root solving; no Gaussian optimization
 assumption enters the converse.}
 \label{fig:allocation}
\end{figure*}

\section{Passive mode mixing and weighted energy}\label{sec:passive}

Consider an $m$-input, $m$-output passive channel
\begin{equation}
 \boldsymbol b=T\boldsymbol a+R\boldsymbol e,~~
 TT^\dagger+RR^\dagger=I_m,
 \label{eq:matrix}
\end{equation}
where the environment is vacuum and $T$ is fixed and known. Its
singular-value decomposition is
\begin{equation}
 T=U\diag(\sqrt{\eta_1},\ldots,\sqrt{\eta_m})V^\dagger .
 \label{eq:svd}
\end{equation}
Changing to input modes $V^\dagger\boldsymbol a$ and output modes
$U^\dagger\boldsymbol b$ gives \eqref{eq:product}.
Vacuum invariance permits the corresponding
environment-mode change. The input and output unitaries can be realized
by passive multiport interferometers \cite{Reck1994,Clements2016}.

\begin{theorem}[Passive multimode channels]\label{thm:passive}
For the channel $\mathcal N_T$ in \eqref{eq:matrix}, with total
input photon number as the cost, all four two-way capacities equal
$\Feta(N)$, where $\eta_j$ are the eigenvalues of $T^\dagger T$.
The same statement holds with a per-use unconditional constraint.
\end{theorem}

\begin{proof}
The input and output transformations in \eqref{eq:svd} are local passive
unitaries. The input transformation preserves
$\sum_ja_j^\dagger a_j$ exactly, for every input state and every
adaptive branch. Both transformations can be absorbed into the
between-use local operations. They give energy-preserving protocol
conversions in both directions, so \cref{thm:capacity} applies.
\end{proof}

This statement concerns multiple modes between one sender and one
receiver. It does not assign a capacity region to a broadcast or
multiple-access network. For a rectangular transfer matrix, vacuum
output modes may be discarded and input null modes have zero capacity;
the same singular-mode formula applies to the remaining modes. Then, we can show the following:

\begin{corollary}[Diagonal energy cost]\label{cor:weighted}
For the multimode pure-loss channel
$\Leta=\bigotimes_{j=1}^{m}\mathcal L_{\eta_j}$
defined in Eq.~\eqref{eq:product}, consider the input cost observable
$H=\sum_jw_j\hat n_j$, with fixed weights $w_j>0$, and impose
$n^{-1}\sum_{i,j}w_jx_{ij}\le E$. Then
\begin{equation}
 \begin{gathered}
 C_2(\Leta,H,E)
 =\max_{\substack{N_j\ge0\\\sum_jw_jN_j\le E}}
       \sum_jc_{\eta_j}(N_j),\\
 C_2\in\{Q_2,D_2,P_2,K_2\}.
 \end{gathered}
 \label{eq:weighted}
\end{equation}
For $E>0$ the optimizer obeys
$c_{\eta_j}'(N_j^*)=\lambda w_j$ and $\sum_jw_jN_j^*=E$.
\end{corollary}

\begin{proof}
The entanglement bound \eqref{eq:adaptivegain} is unchanged. Time
averaging preserves the weighted budget, and the same concavity
argument and product achievability prove \eqref{eq:weighted}.
Strict concavity and the Lagrange multiplier condition give the
optimizer.
\end{proof}

Independent frequency modes have $w_j=\hbar\omega_j$, with the vacuum
energy subtracted. For \eqref{eq:matrix} and
$H=\boldsymbol a^\dagger W\boldsymbol a$, the weighted formula applies
when $W$ and $T^\dagger T$ admit a common eigenbasis, in particular when
$[W,T^\dagger T]=0$. Arbitrary passive mixing preserves total photon
number, but need not preserve a frequency-weighted Hamiltonian. The
case of a noncommuting cost is outside \eqref{eq:weighted}.

\section{Mean-energy strong-converse threshold}\label{sec:strong}

For any of the four tasks, define $C_2^\dagger$ as the infimum of rates
$R$ such that every admissible protocol sequence with
$\liminf_n(\log_2d_n)/n>R$ has error tending to one. For quantum tasks,
this means vanishing entanglement fidelity. The constraint is the
unconditional mean budget in \eqref{eq:energy}. We have the following result:

\begin{theorem}[Unconditional mean constraint]\label{thm:meanstrong}
For every $N>0$ and $0<\eta_j<1$,
\begin{equation}
 \boxed{C_2^\dagger(\Leta,N)=\Beta,}
 \label{eq:meanstrong}
\end{equation}
for $C_2\in\{Q_2,D_2,P_2,K_2\}$.

This also holds under the per-use unconditional constraint. For a square
passive channel with no perfectly transmitting mode,
\begin{equation}
 C_2^\dagger(\mathcal N_T,N)
 =-\log_2\det(I_m-T^\dagger T).
 \label{eq:det}
\end{equation}
\end{theorem}

Appendix~\ref{app:meanstrong} proves the upper bound by tensoring the
comparison states in the unconstrained bosonic converse. The lower
bound uses a public Bernoulli flag: an active protocol of budget
$E>N$ is run with probability $N/E$, and the other branch transmits
vacuum. The rate is defined with the same target size and blocklength
on both branches. The success probability remains bounded away from
zero at any rate below $\Beta$.

\section{Continuous-frequency pure-loss channels}\label{sec:continuum}

\subsection{Spectral-mode operational model}

We now pass from a finite modal list to a frequency profile.
In the spectral regularization, a duration parameter $T$ gives positive
angular frequencies
\begin{equation}
 \omega_k=\frac{2\pi k}{T},~~ k=1,2,\ldots,
 \label{eq:lattice}
\end{equation}
The spacing is $2\pi/T$. Let us call $\mathrm d\omega$ an
infinitesimal interval of angular frequency. In the continuum limit,
$T\,\mathrm d\omega/(2\pi)$ is the number of modes in that interval.
We use one polarization initially. The zero-frequency mode is excluded.
Let the available band $\mathcal B$ be a finite union of compact
intervals in $(0,\infty)$, or all of $(0,\infty)$.
The known transmissivity $\eta:\mathcal B\to[0,1]$ is piecewise
continuous on every compact interval, with finitely many discontinuities
there. Each lattice mode undergoes an independent pure-loss channel
$\mathcal L_{\eta(\omega_k)}$ with vacuum environment.

A finite duration-$T$ protocol selects a finite set of available lattice
modes and uses each selected mode once. Its list and order of calls are
fixed, while local operations, optical inputs, and allocation of energy
may depend on all preceding classical outcomes. Arbitrary local
memories and the public communication resources of Sec.~\ref{sec:model}
are allowed. Vacuum inputs implement branch-dependent nonuse. This
allows interleaving local operations between component calls, and provides a converse for any implementation with fewer opportunities
for feedback. The set of selected modes may grow without bound with $T$.

Fix a mean input-power budget $P\ge0$. For a protocol of optical
duration $T$, the unconditional mean photon numbers $N_{T,k}$
must satisfy
\begin{equation}
 \sum_k\hbar\omega_kN_{T,k}\le PT,
 \label{eq:powerfinite}
\end{equation}
where $\hbar\omega_k$ is the energy of a photon in mode $k$,
and unselected modes have zero mean. Let us call $d_T$ the target
dimension: the Schmidt rank of the maximally entangled state for the
quantum tasks, or the key/message alphabet size for the private tasks.
The rate is $\log_2d_T/T$, and an achievable rate has vanishing
error as $T\to\infty$.
The four resulting rates are denoted
$\mathcal Q_2,\mathcal D_2,\mathcal P_2,\mathcal K_2$.
Two-way classical communication is treated as a free auxiliary
resource. Rates are normalized by the optical transmission duration
$T$, without including additional delays due to classical communication.

\subsection{Capacity and optimal spectral occupation}

Let $\eta(\omega)\in[0,1]$ denote the known channel transmissivity
at angular frequency $\omega\in\mathcal B$. We describe the photon
allocation by a measurable, nonnegative occupation profile
$\nu(\omega)$, representing the mean photon number per spectral mode.
The admissible profiles under a mean input-power budget $P$ form
the set
\begin{equation}
 \mathcal A(P)=\left\{\nu\ge0:
 \int_{\mathcal B}\frac{\mathrm d\omega}{2\pi}
       \hbar\omega\nu(\omega)\le P\right\}.
 \label{eq:profilefeasible}
\end{equation}
Here $\hbar\omega$ is the energy per photon, and
$\mathrm d\omega/(2\pi)$ counts spectral modes per unit time,
so the integral is the total mean input power. The single-mode
capacity contribution is $c_{\eta(\omega)}(\nu(\omega))$, with
$c_0(x)=0$ for complete loss and $c_1(x)=g(x)$ for lossless
transmission. With all this in our hands, we can state the following:

\begin{theorem}[Continuous-frequency capacities]\label{thm:continuum}
In the spectral-mode model above, all four capacity rates coincide:
\begin{equation}
 \mathcal Q_2(P)=\mathcal D_2(P)
 =\mathcal P_2(P)=\mathcal K_2(P)=\Cspec(P),
 \label{eq:continuouscapacities}
\end{equation}
where
\begin{equation}
 \boxed{\Cspec(P)=\sup_{\nu\in\mathcal A(P)}
 \int_{\mathcal B}\frac{\mathrm d\omega}{2\pi}
          c_{\eta(\omega)}(\nu(\omega)).}
 \label{eq:continuouscapacity}
\end{equation}
\end{theorem}

For $P>0$ and transmission over a frequency set of positive measure,
the optimal allocation can be computed using a single parameter
$\lambda>0$. Set $\nu_*(\omega)=0$ where $\eta(\omega)=0$;
at transmitting frequencies, solve
\begin{equation}
 c_{\eta(\omega)}'(\nu_*(\omega))=\lambda\hbar\omega,
 \label{eq:continuouskkt}
\end{equation}
where the prime denotes differentiation with respect to mean photon
number. Choose $\lambda$ so that the full power budget in
Eq.~\eqref{eq:profilefeasible} is used. This condition equalizes the
additional capacity gained per unit energy across transmitting
frequencies. Existence, uniqueness, and the dependence of the capacity
on $P$ are established in Appendix~\ref{app:continuum}.

Note that, for $0<\eta(\omega)<1$, the explicit equation is
\begin{equation}
 \begin{aligned}
 &\log_2\!\left(1+\nu_*^{-1}\right)\\
 &\quad-q(\omega)\log_2\!\left(1+(q(\omega)\nu_*)^{-1}\right)
   =\lambda\hbar\omega,
 \end{aligned}
 \label{eq:continuousslope}
\end{equation}
where $q(\omega)=1-\eta(\omega)$ and the frequency argument of
$\nu_*$ is suppressed. For $\eta=1$ this is the Bose occupation
$\nu_*=(2^{\lambda\hbar\omega}-1)^{-1}$.
The optimization therefore requires only one scalar multiplier,
even for an arbitrary frequency-dependent attenuation profile.

The proof in Appendices~\ref{app:continuum} and~\ref{app:spectralcoding} does not assume
additivity for arbitrary correlated frequency inputs. At every finite
$T$, the replica argument bounds the final entanglement by
$\sum_k c_{\eta(\omega_k)}(N_{T,k})$ for the actual adaptive input
means. The replica limit is completed before increasing $T$.
A scalar dual bound then controls the entire lattice uniformly,
including its low- and high-frequency tails. For achievability,
divide a compact part of the spectrum into narrow intervals, locally
attenuate the outputs in each interval to a common transmissivity,
and apply the finite-dimensional sector-hashing codes. Increasing
$T$ supplies arbitrarily many modes in each interval. Refining the
intervals and exhausting the spectrum gives \eqref{eq:continuouscapacity}.

One useful consequence is the universal bound
\begin{equation}
 \Cspec(P)\le\mathcal R_{\mathrm{id}}(P)
 :=\frac1{\ln2}\sqrt{\frac{\pi P}{3\hbar}}.
 \label{eq:idealrate}
\end{equation}
Thus the frequency-weighted energy constraint makes the infinite-band
rate finite. A photon-number budget without the factor $\hbar\omega$
would describe a different optimization and would not supply this
high-frequency control.

\subsection{Frequency-preserving mode mixing}

A finite number $d$ of polarizations or spatial channels per frequency
is included by summing the integrand and the power cost over their
index. More generally, let a known matrix $\mathsf T(\omega)$ describe
passive mixing at each frequency. Its singular-mode transformation
acts within the eigenspace of the energy $\hbar\omega$ and therefore
preserves the Hamiltonian. If $\eta_a(\omega)$ are its squared
singular values and are locally piecewise continuous, the capacity is
\begin{equation}
 \sup_{\{\nu_a\}}
 \sum_{a=1}^{d}\int_{\mathcal B}\frac{\mathrm d\omega}{2\pi}
       c_{\eta_a(\omega)}(\nu_a(\omega)),
 \label{eq:matrixprofile}
\end{equation}
subject to the corresponding total power constraint.
Each occupied mode satisfies
$c_{\eta_a(\omega)}'(\nu_a)=\lambda\hbar\omega$.
This follows by applying the finite-mode passive reduction and the
same continuum proof; the integrable envelope is multiplied by $d$.
Mixing different physical frequencies need not preserve the energy
cost and is not covered by this reduction.

\section{Broadband channel with constant transmissivity}\label{sec:broadband}

For a frequency-independent $\eta$ and $\mathcal B=(0,\infty)$,
the variational problem admits a closed expression. Define
\begin{equation}
 A_\eta=\eta\Li_2(\eta)
          -\frac{1-\eta}{2}\ln^2(1-\eta),
 \label{eq:Aeta}
\end{equation}
where $\Li_2$ is the dilogarithm, defined on $0\le z\le1$ by
$\Li_2(z)=\sum_{r=1}^{\infty}z^r/r^2$ \cite{DLMF}.
In particular, $\Li_2(0)=0$ and $\Li_2(1)=\pi^2/6$.
The coefficient has the continuous endpoint values $A_0=0$ and
$A_1=\pi^2/6$, with $(1-\eta)\ln^2(1-\eta)$ understood by its limit
at $\eta=1$. Given these preliminary definitions, we can state the following result:

\begin{theorem}[Exact broadband square-root law]\label{thm:broadband}
For constant $0\le\eta\le1$, one polarization, and mean input power $P$,
all four capacity rates in \cref{thm:continuum} equal
\begin{equation}
 \boxed{\mathcal C_\eta^{\mathrm{bb}}(P)
       =\frac1{\ln2}\sqrt{\frac{2A_\eta P}{\pi\hbar}}.}
 \label{eq:broadband}
\end{equation}
For $0<\eta<1$, put $q=1-\eta$ and define $v_\eta(x)$ by
\begin{equation}
 \ln(1+v_\eta^{-1})
 -q\ln(1+(qv_\eta)^{-1})=x.
 \label{eq:veta}
\end{equation}
The optimal spectrum is
\begin{equation}
 \nu_*(\omega)=v_\eta(\omega/\Omega_\eta),~~
 \Omega_\eta=\sqrt{\frac{2\pi P}{\hbar A_\eta}}.
 \label{eq:broadbandspectrum}
\end{equation}
For $\eta=1$, use $v_1(x)=(e^x-1)^{-1}$.
\end{theorem}

The derivation in Appendix~\ref{app:broadband} evaluates the identity
\begin{equation}
 \begin{aligned}
 A_\eta&=\int_0^\infty x v_\eta(x)\,\mathrm dx\\
 &=\frac12\int_0^\infty
 \left[\ln(1+u^{-1})-q\ln(1+(qu)^{-1})\right]^2\mathrm du.
 \end{aligned}
 \label{eq:Aintegral}
\end{equation}
This evaluates the square-root coefficient for the two-way capacity
obtained from the single-mode rate $c_\eta$.

\begin{figure*}[t]
\centering
\includegraphics[width=\textwidth]{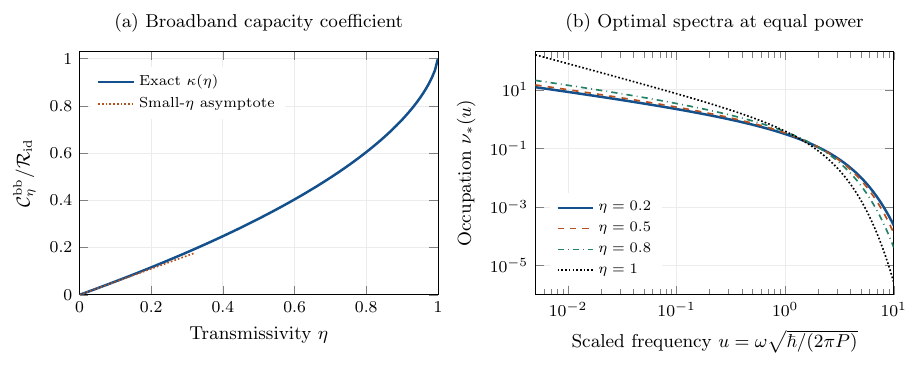}
\caption{Broadband two-way capacities and optimal spectra.
(a) The exact coefficient $\kappa(\eta)$ in \eqref{eq:kappa}; the
dotted segment is its small-$\eta$ asymptote. (b) Optimal occupations
at the same total power for four constant transmissivities. The
dimensionless frequency is $u=\omega\sqrt{\hbar/(2\pi P)}$, so every
curve satisfies $\int_0^\infty u\nu_*(u)\,\mathrm du=1$.
For every strictly lossy channel the low-frequency divergence is
$u^{-1/2}$, whereas the ideal channel has a $u^{-1}$ occupation.
All spectra have exponentially decreasing high-frequency tails.}
\label{fig:broadband}
\end{figure*}

Relative to the lossless rate in \eqref{eq:idealrate}, write
\begin{equation}
 \mathcal C_\eta^{\mathrm{bb}}(P)
 =\kappa(\eta)\mathcal R_{\mathrm{id}}(P),~~
 \kappa(\eta)=\frac{\sqrt{6A_\eta}}{\pi}.
 \label{eq:kappa}
\end{equation}
The factor is strictly increasing, has endpoints zero and one,
and obeys
\begin{equation}
 \kappa(\eta)=\frac{\sqrt3}{\pi}\eta+O(\eta^2),~~
 \kappa(1/2)\simeq0.322428.
 \label{eq:kappachecks}
\end{equation}
Figure~\ref{fig:broadband} shows this factor and the optimizing spectra.
For $d$ identical independent polarizations or spatial channels,
sharing a total power $P$, concavity gives equal powers $P/d$ and
multiplies \eqref{eq:broadband} by $\sqrt d$.

\section{Mean-power strong-converse thresholds}\label{sec:continuousstrong}

Define $\mathcal C_2^\dagger(P)$ by replacing blocklength with $T$
in the strong-converse definition of \cref{sec:strong}. Let
\begin{equation}
 b(\eta)=-\log_2(1-\eta),~~
 \Bspec=\int_{\mathcal B}\frac{\mathrm d\omega}{2\pi}
                       b(\eta(\omega)),
 \label{eq:Bspec}
\end{equation}
with $b(0)=0$ and $b(1)=+\infty$. Then, we have:

\begin{theorem}[Spectral mean-power threshold]\label{thm:continuousstrong}
Let $P>0$. If $\Bspec=+\infty$, then
$\mathcal C_2^\dagger(P)=+\infty$ for each of the four tasks.
If $\Bspec<\infty$, assume that no available lattice mode is ideal
and that the spectral trace condition
\begin{equation}
 \frac1T\sum_{k:\,\omega_k\in\mathcal B}b(\eta(\omega_k))
       \longrightarrow\Bspec
 \label{eq:spectraltrace}
\end{equation}
holds. The sum runs over all available lattice frequencies:
it has finitely many terms on a compact band, whereas on an unbounded
band the nonnegative series must converge to a finite value for every
sufficiently large $T$. Then
\begin{equation}
 \boxed{\mathcal C_2^\dagger(P)=\Bspec.}
 \label{eq:continuousstrong}
\end{equation}
\end{theorem}

For a compact band, \eqref{eq:spectraltrace} holds whenever the
transmissivity is bounded away from one. On the full positive spectrum,
a sufficient condition is that $b(\eta(\omega))$ is dominated by a
nonnegative, nonincreasing, integrable function. These conditions
control the unconstrained rate density, not merely the finite-power
capacity integrand. They are additional to the assumptions of
\cref{thm:continuum}.

The proof in Appendix~\ref{app:continuousstrong} combines the finite-list
unconstrained converse with the trace limit. Conversely,
$\Cspec(E)\uparrow\Bspec$ as $E\to\infty$.
Running an active power-$E$ code with public probability $P/E$
preserves the mean power and a nonzero success probability.
In particular, the idealized unlimited-bandwidth channel with
constant $\eta>0$ has
\begin{equation}
 \mathcal C_\eta^{\mathrm{bb}}(P)<\infty,
 ~~ \mathcal C_2^\dagger(P)=+\infty.
 \label{eq:bbthreshold}
\end{equation}
The active branch uses a finite but arbitrarily large power; no uniform
bound on its power is imposed by an unconditional mean constraint.

\section{Consequences for multimode and broadband loss}\label{sec:multimodeconsequences}

For a device with calibrated transmissivities, the allocation problem
quantifies the rate sacrificed by a prescribed photon distribution.
For the three-mode example $\boldsymbol\eta=(0.8,0.4,0.1)$ and $N=3$,
the optimizer is
\begin{equation}
 \begin{gathered}
 \boldsymbol N^*\simeq(2.128494,\,0.758918,\,0.112588),\\
 \Feta(3)\simeq2.042441.
 \end{gathered}
 \label{eq:example}
\end{equation}
Figure~\ref{fig:allocation} compares this optimum with simpler feasible
allocations over a range of budgets.

The rate above is per three-mode use. In the continuum, the rate is
per unit time and the budget is a power. The mode density
$T\,\mathrm d\omega/(2\pi)$ and photon energy $\hbar\omega$
specify the cost of accessing additional frequencies. An unweighted
photon budget would therefore define a different broadband problem.

A flat transmission spectrum provides a reference for the spectral
optimization. Frequency-dependent loss introduces additional scales,
so the power dependence need not follow the same square-root law.
For a given profile, Eq.~\eqref{eq:continuouskkt} can be evaluated
on frequency bins, with the spectral tail estimates controlling the
contribution omitted outside the numerical window.

The distinction between capacity and strong-converse threshold also
matters when interpreting a power specification. An unconditional mean
bound permits rare energetic branches. For unlimited bandwidth and
constant nonzero loss, the finite-power capacity remains finite while
the threshold is infinite; for a decaying profile, the trace condition
can instead give a finite threshold. These are statements about
averaging over all protocol outcomes, rather than peak-power guarantees.

The continuum theorem uses independently addressable spectral modes.
Relating this model to a finite-time optical scattering problem requires
control of temporal boundary coupling. Similarly, a transformation
that mixes physical frequencies need not preserve the power cost.
These restrictions delimit the direct application of the spectral
formula to an optical device.

\section{Fading models and channel-state information}
\label{fd:sec:models}
We now turn from known deterministic loss profiles to a randomly
selected attenuator. Terrestrial entanglement distribution
\cite{Ursin2007} and satellite demonstrations of key distribution
\cite{Liao2017} and entanglement distribution \cite{Yin2017}
provide experimental context for this model.
The entropy and pure-loss rate functions remain
those of \cref{fd:eq:g,fd:eq:c}, and the operational resources and
error conventions are those of \cref{sec:model}.
Let \(\mu\) be a Borel probability measure on \([0,1]\). Transmissivities
\(\eta_1,\eta_2,\ldots\) are drawn independently with law \(\mu\).
There is no additional thermal noise or random phase rotation.
We use the moments
\begin{equation}
\begin{aligned}
\bar\eta&=\int\eta\,\mu(\dd\eta),~~
a=\int\sqrt{\eta}\,\mu(\dd\eta),\\
\beta&=a^2.
\end{aligned}
\label{fd:eq:moments}
\end{equation}
Thus \(0\leq\beta\leq\bar\eta\leq1\).
All entropies are in bits, and \(h_2\) denotes binary entropy.

Each protocol uses \(n\) optical transmission slots. Arbitrary local
quantum processing, quantum storage, and authenticated public
communication may be interleaved with these transmissions. The initial
laboratories contain no shared entanglement or secret key; public
randomness is allowed. Only the transmitted optical modes are charged
to the energy budget. Receiver memories, local ancillary systems, and
the shields of private states have no energy restriction and may act
on arbitrary separable Hilbert spaces. Eve receives a purification of
the channel noise and the public transcript.

For the unconditional input state \(\rho^{(i)}_{A_i}\), the constraint is
\begin{equation}
\frac1n\sum_{i=1}^{n}\Tr(\hat n_{A_i}\rho^{(i)}_{A_i})\leq N.
\label{fd:eq:budget}
\end{equation}
This expectation includes every public branch and every fading
realization. A vacuum input, an unsuccessful transmission, and a
discarded output still occupy a channel slot. Neither a peak-energy
constraint nor a budget conditioned on successful transmission is
imposed.

The two-way quantum and entanglement-distribution capacities are
denoted by \(Q_2\) and \(D_2\), and the secret-key capacity by \(K_2\).
Rates are normalized by \(n\), and the error tends to zero.
Quantum targets are maximally entangled states of dimension \(d_n\).
For key agreement one may retain local purifications and use an
approximate private state with key dimension \(d_n\)
\cite{HHHO2005,Horodecki2009}. Thus the relevant rate is
\(\liminf_n n^{-1}\log_2d_n\). Local teleportation converts distributed
entanglement into quantum transmission without changing
Eq.~\eqref{fd:eq:budget}, so \(Q_2=D_2\leq K_2\)
\cite{Teleportation1993}. The two-way private-message capacity \(P_2\) equals
\(K_2\) by using a one-time pad and public ciphertext.

\subsection{Receiver information}
The current \(\eta_i\) is disclosed publicly only after the input has
been chosen. A law has \emph{finite support} if the transmissivity
can take only finitely many values $\eta_1,\ldots,\eta_d$, with
probabilities $w_j>0$ satisfying $\sum_jw_j=1$. We write
$\mu=\sum_{j=1}^{d}w_j\delta_{\eta_j}$, where $\delta_{\eta_j}$
is the Dirac probability measure concentrated at $\eta_j$:
$\int f(\eta)\,\delta_{\eta_j}(\dd\eta)=f(\eta_j)$.
For this law, the channel with its receiver record is
\begin{equation}
\mathcal F_{\rm rec}(\rho)
=\sum_j w_j\mathcal L_{\eta_j}(\rho)\otimes\proj j_F.
\label{fd:eq:receiverchannel}
\end{equation}
Bob can copy this classical label to Alice. The input at use \(i\)
may depend on all previous records, but is independent of the current
draw. General Borel records are treated by classical conditioning and
finite quantization; no entropy is assigned to a continuous label.

A physical motivation is a link monitored by additional classical
pilot pulses. If pointing errors and turbulence evolve slowly compared
with the detector time resolution and the spacing of suitable pilots,
Bob can estimate the loss experienced by nearby quantum signals and
announce it after transmission
\cite{Pirandola2021,PirandolaSatellite2021,PirandolaTrust2021}.
The receiver-information model idealizes this estimate as an exact
public record. Its budget charges the quantum inputs; the auxiliary
pilots and their estimation errors are not modeled. The independent
draws assumed here describe the selected channel uses. Correlations
between successive uses require a channel model with memory, for
example block fading, in which each group of uses shares one
transmissivity.

\subsection{Advance information}
The current \(\eta_i\) is supplied publicly before Alice chooses the
input at that use. The conditional input energy
\(x_i(\eta)=\Tr(\hat n\rho^{(i,\eta)}_{A_i})\) may therefore depend on
\(\eta\), and the budget becomes
\begin{equation}
\frac1n\sum_i\int x_i(\eta)\,\mu(\dd\eta)\leq N.
\label{fd:eq:prebudget}
\end{equation}
The realization is side information, not a signal that Alice can choose
or modulate. All fading states, including complete blockage, count
towards the rate.

\subsection{Unobserved fading}
No realization is disclosed to either party. The memoryless channel is
\begin{equation}
\F(\rho)=\int\mathcal L_\eta(\rho)\,\mu(\dd\eta).
\label{fd:eq:hiddenchannel}
\end{equation}
Eve is assigned a complementary channel of this averaged map.
This model is relevant when loss fluctuations are not resolved by the
available receiver and channel tracking, for example when their
dynamics are fast compared with the effective detection or estimation
time \cite{Pirandola2021,Papanastasiou2018}. An implementation may
then operate without separate tracking pilots, processing the averaged
output directly. Such pilots are not required to define or use this
model. More generally, detector averaging must be checked against the
chosen optical modes; it does not by itself imply the particular
memoryless mixture in Eq.~\eqref{fd:eq:hiddenchannel}.

\section{Exact capacities with fading information}
\label{fd:sec:observed}
\begin{theorem}[Receiver information]
\label{fd:thm:receiver}
For every finite \(N\geq0\), the receiver-information model satisfies
\begin{equation}
\boxed{\begin{aligned}
Q_2=D_2=P_2=K_2&=\Crec(N),\\
\Crec(N)&=\int c_\eta(N)\,\mu(\dd\eta).
\end{aligned}}
\label{fd:eq:receivercapacity}
\end{equation}
\end{theorem}

The same input preparation is optimal for every conditional pure-loss
channel: one half of a two-mode squeezed vacuum with mean transmitted
energy \(N\). Reverse hashing, conditioned on the publicly shared
record, achieves Eq.~\eqref{fd:eq:receivercapacity}
\cite{Reverse,DirectReverse,DevetakWinter}.
The converse applies the pure-loss sector bound to independent
replicas of the adaptive protocol, grouped by the current fading
value. This produces the average of \(c_\eta(x_i)\), where \(x_i\)
does not depend on the current realization. Its concavity gives
\(\sum_i\Crec(x_i)\leq n\Crec(N)\). Appendix~\ref{fd:app:observed}
supplies the full argument, including continuous fading.

\begin{theorem}[Advance information]
\label{fd:thm:advance}
For the advance-information model,
\begin{equation}
\boxed{\begin{aligned}
Q_2=D_2=P_2=K_2&=\Cpre(N),\\
\Cpre(N)&=\sup_{\int x(\eta)\mu(\dd\eta)\leq N}
\int c_\eta(x(\eta))\,\mu(\dd\eta).
\end{aligned}}
\label{fd:eq:precapacity}
\end{equation}
where the optimization is over nonnegative measurable allocations.
If \(\mu((0,1])>0\) and \(N>0\), an optimal allocation exhausts the
budget and is unique almost everywhere on \(\eta>0\).
\end{theorem}

Set \(x_\lambda(0)=0\). For \(\eta>0\), the optimal allocation is the
unique positive solution of
\begin{align}
\partial_xc_\eta(x_\lambda(\eta))
&=\lambda,\label{fd:eq:allocation}\\
\partial_xc_\eta(x)
&=\log_2(1+x^{-1})\nonumber\\
&\quad-(1-\eta)\log_2\!\left(1+\frac1{(1-\eta)x}\right),\nonumber
\end{align}
where the second term is zero at \(\eta=1\). The multiplier
\(\lambda>0\) is chosen so that \(\int x_\lambda\,\dd\mu=N\).
The allocation increases with transmissivity. With only the mean
constraint, it is positive for every \(\eta>0\), although it may be
arbitrarily small in poor fading states. Appendix~\ref{fd:app:allocation}
proves these assertions.

Discarding information and choosing a constant allocation give
\begin{equation}
Q_2(\F,N)\leq K_2(\F,N)\leq\Crec(N)\leq\Cpre(N)\leq g(N).
\label{fd:eq:hierarchy}
\end{equation}
Receiver information does not permit selection of the current input
energy: past records are independent of the next draw. In contrast,
Eq.~\eqref{fd:eq:precapacity} uses information available at the time the
input is selected.

\section{Finite-energy bounds for unobserved fading}
\label{fd:sec:comparison}
The bounds for the averaged channel can be expressed through a single
variational quantity. Let $\pi=(\pi_n)_{n\ge0}$ be a photon-number
distribution. For an input number $n$ and a loss count
$0\le\ell\le n$, define
\begin{equation*}
 v_{n\ell}(\eta)=\sqrt{\binom n\ell}\,
 \eta^{(n-\ell)/2}(1-\eta)^{\ell/2},
\end{equation*}
with continuous endpoint values and $v_{n\ell}=0$ for $\ell>n$.
These functions are vectors in $L^2(\mu)$, the space of functions
square-integrable with respect to the fading law $\mu$. Placing
different loss counts in orthogonal subspaces gives the auxiliary states
\begin{equation*}
 \tau_n=\bigoplus_{\ell\ge0}\proj{v_{n\ell}}
 \quad\text{on}\quad
 \bigoplus_{\ell\ge0}L^2(\mu).
\end{equation*}
Each $\tau_n$ has unit trace because
$\sum_\ell\int|v_{n\ell}(\eta)|^2\mu(\dd\eta)=1$.
Write $H(\pi)=-\sum_n\pi_n\log_2\pi_n$ for Shannon entropy and
$S(\rho)=-\Tr(\rho\log_2\rho)$ for von Neumann entropy, both in
bits \cite{Simeone2026}. Define
\begin{equation*}
 \begin{aligned}
 \Afun(\pi)={}&H(\pi)+\sum_n\pi_n S(\tau_n)\\
 &-S\!\left(\sum_n\pi_n\tau_n\right).
 \end{aligned}
\end{equation*}
This conditional entropy measures the uncertainty about the input
photon number remaining when the corresponding auxiliary state is
available. For a mean photon-number budget $N$, set
\begin{equation}
 U_\mu(N)=\sup_{\substack{\pi_n\ge0,\ \sum_n\pi_n=1\\
                         \sum_n n\pi_n\le N}}\Afun(\pi).
 \label{fd:eq:U}
\end{equation}
The functional is nonnegative and concave, and
$0\le\Afun(\pi)\le H(\pi)\le g(N)$ at mean at most $N$.
Appendix~\ref{fd:app:extension} establishes these properties and
constructs the comparison channel used in the proof. The on/off
example below reduces the optimization to one scalar. With the definitions introduced above, we can finally state the following: 

\begin{theorem}[Finite-energy bounds for unobserved fading]
\label{fd:thm:extension}
For the unobserved channel $\F$ under Eq.~\eqref{fd:eq:budget},
\begin{multline}
\max\{0,U_\mu(N)-Nh_2(\bar\eta)\}
\leq Q_2(\F,N)=D_2(\F,N)\\
\leq K_2(\F,N)
\leq\min\{U_\mu(N),\Crec(N)\}.
\label{fd:eq:bracket}
\end{multline}
Consequently,
\begin{equation}
0\leq K_2(\F,N)-Q_2(\F,N)\leq Nh_2(\bar\eta).
\label{fd:eq:keygap}
\end{equation}
\end{theorem}

Appendix~\ref{fd:app:extension} proves the bounds using an auxiliary
loss-count record for the converse and reverse hashing for
achievability. An additional single-photon rail bound is derived in
Appendix~\ref{fd:app:rail} and included in the numerical comparison
below. For $\bar\eta=0$, all capacities and bounds vanish.

\section{Examples: binary and on/off fading}
\label{fd:sec:binary}
For \(\mu=p\delta_{\eta_0}+(1-p)\delta_0\), with
\(p,\eta_0>0\), Theorems~\ref{fd:thm:receiver} and
\ref{fd:thm:advance} give
\begin{align}
\Crec(N)&=p\,c_{\eta_0}(N),\label{fd:eq:binaryrec}\\
\Cpre(N)&=p\,c_{\eta_0}(N/p).\label{fd:eq:binarypre}
\end{align}
At \(p=0\), all rates vanish. For the on/off channel
\begin{equation}
\Phi_p=p\,\id+(1-p)\mathcal L_0,
~~ 0<p<1,
\label{fd:eq:onoff}
\end{equation}
the upper-bound and reverse-hashing optimizations reduce exactly to
one scalar. Define
\begin{align}
t_\epsilon&=1-(1-p)\epsilon,\nonumber\\
s_\epsilon&=\frac{t_\epsilon-
\sqrt{t_\epsilon^2-4p(1-p)\epsilon(1-\epsilon)}}2,\nonumber\\
j_p(\epsilon)&=h_2(\epsilon)-
H(t_\epsilon-s_\epsilon,s_\epsilon,(1-p)\epsilon).
\label{fd:eq:binaryj}
\end{align}
Writing \(R_p(N)\) for optimized reverse coherent information,
we obtain
\begin{align}
U_p(N)&=\max_{0\leq\epsilon\leq\min\{N,1\}}
\left[j_p(\epsilon)+\epsilon h_2(p)
+p\epsilon g(N/\epsilon-1)\right],\label{fd:eq:onoffU}\\
R_p(N)&=\max_{0\leq\epsilon\leq\min\{N,1\}}
\left[j_p(\epsilon)+p\epsilon g(N/\epsilon-1)\right].
\label{fd:eq:onoffR}
\end{align}
Both objectives are assigned their continuous value zero at
\(\epsilon=0\). The maximizing conditional distribution on
\(n\geq1\) is geometric after shifting photon number by one.
Appendix~\ref{fd:app:binary} derives these formulas and proves that
Fock-diagonal inputs suffice for reverse coherent information.
Neither \(U_p\) nor \(R_p\) is identified with the full unobserved
capacity.

\begin{figure}[!tbp]
\centering
\includegraphics[width=\columnwidth]{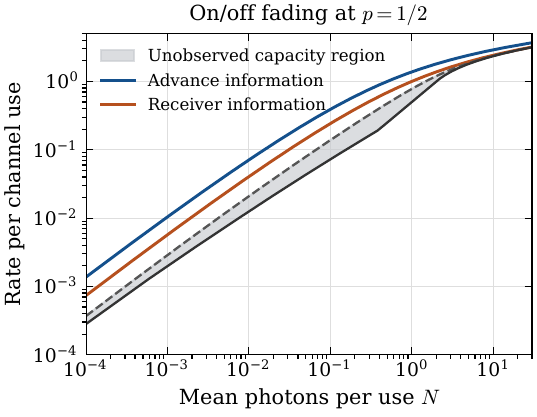}
\caption{Role of fading information for
\(\Phi_p=p\,\id+(1-p)\mathcal L_0\) at \(p=1/2\).
The exact advance-information (blue) and receiver-information
(orange) capacities are compared with bounds on every unobserved
assisted capacity (gray). The dashed edge is a converse; the solid
edge is achievable. All rates count every optical channel use,
including blocked and vacuum slots.}
\label{fd:fig:capacities}
\end{figure}

Figure~\ref{fd:fig:capacities} compares the two exact observed capacities
with bounds on the unobserved capacities. The shaded region is
bounded above by \(\min\{U_p(N),p g(N)\}\).
Its lower edge is the largest of reverse hashing, the evaluated
single-photon rail rates, and time-shared erasure coding on the
strictly positive-photon subspace.

For \(p=1/2\) and \(N=0.1\), the bounds give
\begin{align}
\Cpre&\simeq0.390013,~~ \Crec\simeq0.241723,\nonumber\\
0.072384&\lesssim Q_2=D_2\leq K_2\lesssim0.138232.
\label{fd:eq:numerical}
\end{align}

\section{Mean-energy strong-converse thresholds for fading}
\label{fd:sec:strong}

The strong-converse threshold of \cref{sec:strong} also extends to
fading. It must be distinguished from the vanishing-error capacity at
the specified finite mean energy. Define the extended average
\begin{equation}
 \Bfade=\int_{[0,1]}b(\eta)\,\mu(\dd\eta),
 ~~ b(\eta)=-\log_2(1-\eta),
 \label{fd:eq:strongaverage}
\end{equation}
with $b(0)=0$ and $b(1)=+\infty$. The fading draws remain independent
and identically distributed, and every transmission slot is counted. We have the following:

\begin{theorem}[Mean-energy fading thresholds]\label{fd:thm:strong}
For every $N>0$, each of the four tasks has the following
strong-converse thresholds under the unconditional mean constraint.
For receiver and advance information, and any Borel fading law,
\begin{equation}
 \boxed{\Crec^\dagger(N)=\Cpre^\dagger(N)=\Bfade,}
 \label{fd:eq:observedstrong}
\end{equation}
including the value $+\infty$.
For unobserved fading with finite support,
\begin{equation}
 \boxed{C_2^\dagger(\F,N)=\Bfade,
 ~~ C_2\in\{Q_2,D_2,P_2,K_2\}.}
 \label{fd:eq:hiddenstrong}
\end{equation}
For an arbitrary unobserved Borel law with $\Bfade<\infty$,
\begin{equation}
 C_2^\dagger(\F,N)\leq\Bfade.
 \label{fd:eq:generalstrongupper}
\end{equation}
At $N=0$, all thresholds are zero.
\end{theorem}

The finite-support assumption means
$\mu=\sum_{j=1}^{d}w_j\delta_{\eta_j}$ with a fixed finite number
$d$ of transmissivities, positive probabilities $w_j$, and
$\sum_jw_j=1$. It restricts the fading law while leaving the optical
Hilbert space infinite dimensional. This assumption is needed for
the high-energy coding result used to obtain equality for unobserved
fading. For a general unobserved continuous law, even on a bounded interval,
only the stated upper bound is established here. The proof is given in Appendix~\ref{fd:app:strong}.

For binary fading $\mu=p\delta_{\eta_0}+(1-p)\delta_0$, the three
information regimes therefore share
\begin{equation}
 C_2^\dagger(N)=-p\log_2(1-\eta_0),
 ~~ N>0,
 \label{fd:eq:binarystrong}
\end{equation}
when $0<\eta_0<1$. For the on/off channel with $\eta_0=1$ and
$p>0$, the threshold is infinite, although every finite-energy
capacity is finite. In that case the erasure-code rate of
Appendix~\ref{fd:app:binary} already diverges with the active-branch energy,
so the infinite threshold also follows directly from that construction.

\section{Discussion of fading}\label{sec:discussion}
The public-record models describe what can be gained by resolving loss
fluctuations. Their ideal side information must be related to channel
tracking and detector time resolution in a concrete implementation.
A noisy or partially informative record defines a different channel
resource, whose value depends on both its accuracy and when it
becomes available.

For unobserved fading, the present bounds do not settle whether
entanglement distribution and secret-key agreement have the same
finite-energy capacity. In particular, Eq.~\eqref{fd:eq:keygap}
controls their possible separation without proving equality at a
fixed budget. Binary on/off fading is a useful test case for improving
distillation protocols and comparing them with the converse.
For a continuous fading law, a possible route to the unresolved lower
strong-converse threshold is to control the energy cost of identifying
increasingly close attenuation values.

Independence between successive draws is also essential to the stated
information regimes: previous records do not predict the next
transmissivity. In block fading, measurements early in a coherence
interval can inform later inputs, so the estimation and coding stages
must be treated together. Thermal background noise and random phase
introduce further channel parameters that a distribution of pure-loss
transmissivities alone does not describe.

\section{Conclusions}\label{sec:conclusions}
We have extended the finite-energy pure-loss converse from
Ref.~\cite{Pirandola2026} to arbitrary adaptive multimode \mbox{protocols}
and continuous frequencies. The deterministic results give
exact two-way quantum and secret-key capacities as optimized modal
sums or spectral integrals. For constant loss over an unlimited band,
the integral has a closed form proportional to the square root of
input power.

For passive multimode links, the squared singular values of the
transmission matrix determine the capacity under a total photon-number
constraint. The optimal allocation is unique on the transmitting
modes: equal losses receive equal occupations, and more transmissive
modes receive more photons. A common marginal information gain per unit
cost also determines weighted allocations when the energy observable is diagonal
in the loss basis. Independent inputs in that basis attain the
capacity of the full adaptive protocol class.

For fading, we obtained exact capacities in both observed models and
computable bounds for the unobserved mixture, including scalar
expressions for on/off loss. We also identified the mean-constrained
strong-converse thresholds for deterministic loss under the stated
spectral conditions and for observed fading. For unobserved fading,
the threshold is exact for finite-support laws, while a general law
with finite average PLOB rate has the corresponding upper bound.

A natural next step is a joint model in which a multimode transmission
matrix varies in time, linking spatial and spectral allocation to
partial channel-state information. Finite-blocklength versions of the
present bounds could also guide the choice of coding duration and
optical bandwidth at a prescribed error. These directions would connect
the asymptotic allocation formulas more directly to the resources of
an implemented communication system.

\section*{AI Statement}
OpenAI's ChatGPT (GPT-6 Astra model) was used during the
preparation of this manuscript to generate candidate mathematical
derivations and proofs, assist with their development, and improve the presentation of the manuscript. The tool was directed through
author-specified prompts concerning the mathematical assumptions,
derivations, and statements under consideration. All AI-assisted
mathematical results, derivations, and proofs were independently
examined, verified, and revised as necessary by the authors, who assume responsibility for the accuracy and scientific content of the work.

\appendix

\section{Structure of the proof and sector tools}\label{app:structure}

\begin{figure*}[t]
\centering
\includegraphics[width=0.94\textwidth]{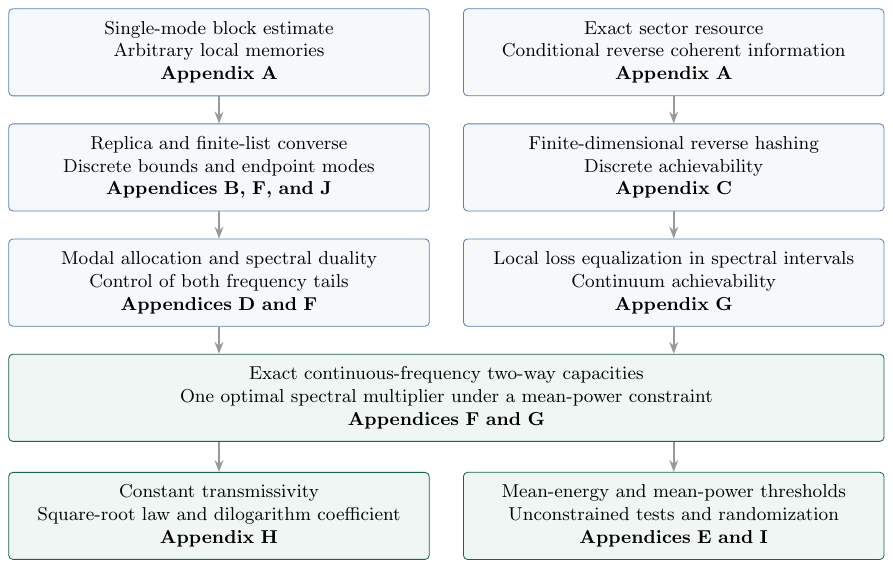}
\caption{Structure of the proof. The single-mode block estimate
(\cref{lem:importedblock}) yields the adaptive replica and finite-list
bounds (\cref{lem:replica,lem:finitelist}). Scalar duality and spectral
tail estimates turn these into the continuum converse. The sector
resource (\cref{lem:sector}) supplies finite-dimensional hashing,
first for discrete modes and then for spectral intervals. Together
these branches prove \cref{thm:continuum}. Evaluating the flat-profile
integral gives \cref{thm:broadband}. The threshold statements also use
the unconstrained resource comparison and public vacuum randomization,
as detailed in Appendices~\ref{app:meanstrong} and~\ref{app:continuousstrong}.}
\label{fig:proof}
\end{figure*}

For the deterministic multimode and broadband results, the
channel-specific input is the single-mode sector estimate stated below. We keep the number of physical
modes $m$, the communication blocklength $n$, and the replica number $r$
distinct. The replica limit is always taken with the complete finite
protocol fixed. The duration $T$ of the spectral regularization is
a separate parameter, and its limit is taken only after the
finite-protocol replica bound has been obtained.
Figure~\ref{fig:proof} summarizes the discrete and continuous arguments
and identifies the appendices containing each step.
The fading proof structure is given separately in Appendix~\ref{fd:app:map};
its observed models reuse the same single-mode block estimate.

For a state $\rho_{AB}$, define
\begin{equation}
 \begin{gathered}
 \ER(A:B)_\rho=\inf_{\sigma\in\SEP(A:B)}D(\rho\Vert\sigma),\\
 \ERi(\rho)=\lim_{r\to\infty}\frac1r\ER(\rho^{\otimes r}),
 \end{gathered}
 \label{eq:ree}
\end{equation}
where the separable set is trace-norm closed and
$D(\rho\Vert\sigma)=\Tr\rho(\log_2\rho-\log_2\sigma)$, with the
usual support convention. The regularization exists whenever the
single-copy REE is finite, by subadditivity. We use LOCC
monotonicity and subadditivity, not additivity of REE.

\begin{lemma}[Single-mode block gain]\label{lem:importedblock}
Let $\rho_{XA^sY}$ have finite $\ER(XA^s:Y)_\rho$ and mean
transmitted photon number
$\Tr[(\sum_{u=1}^s\hat n_{A_u})\rho]=sx<\infty$. For
$\omega=(\id_{XY}\otimes\mathcal L_\eta^{\otimes s})(\rho)$,
\begin{equation}
 \begin{aligned}
 \ER(X:B^sY)_\omega
 \le{}&\ER(XA^s:Y)_\rho+s c_\eta(x)\\
 &+g(s\eta x)+g(s(1-\eta)x).
 \end{aligned}
 \label{eq:importedblock}
\end{equation}
The memories may be correlated, entangled, and infinite dimensional.
\end{lemma}

This is the block entanglement-gain lemma of
Ref.~\cite{Pirandola2026}, Appendix B. Its proof combines sector
teleportation with control of number-sector coherence. The two count
entropies in \eqref{eq:importedblock} are $o(s)$ at fixed finite $x$.
It is important that the lemma concerns an arbitrary joint input,
rather than a product-state optimization.

We also record the sector tool used in the achievability arguments. Let $\mathcal H_{s,k}$ be the $k$-photon subspace of $s$ modes,
with projector $P_{s,k}$ and dimension
\begin{equation}
 d_{s,k}=\binom{s+k-1}{k}.
 \label{eq:dimension}
\end{equation}
Write $\Phi_{s,k}$ for a maximally entangled state between this subspace
and an isomorphic reference, and define
\begin{equation}
 J^{(\eta)}_{s,k}
   =(\id\otimes\mathcal L_\eta^{\otimes s})(\Phi_{s,k}).
\end{equation}

\begin{lemma}[Sector simulation and hashing]\label{lem:sector}
There is a deterministic LOCC map $\mathcal T^{(\eta)}_{s,k}$ such that
\begin{equation}
 \begin{gathered}
 (\id_{XY}\otimes\mathcal L_\eta^{\otimes s})(\rho)
 =\mathcal T^{(\eta)}_{s,k}
       \big(\rho\otimes J^{(\eta)}_{s,k}\big),\\
 \supp\rho_{A^s}\subseteq\mathcal H_{s,k}.
 \end{gathered}
 \label{eq:sectorsim}
\end{equation}
The identity includes arbitrary reference and memory correlations.
The resource has a locally flagged decomposition by the lost number
$L\sim\Bin(k,1-\eta)$. Its normalized conditional state
$J^{(\eta)}_{s,k\mid\ell}$ has reverse coherent information
$\log_2d_{s,k}-\log_2d_{s,\ell}$.
\end{lemma}

These are the sector simulation and conditional-resource statements of
Ref.~\cite{Pirandola2026}, Appendices B and G. Covariance acts on
the $s$ identical-transmissivity modes, whose fixed-number
representation is irreducible. In the present application, such modes
are either replicas of one transmission or the temporal repetitions of
one physical mode, or frequency modes whose effective losses have
been equalized locally. No covariance under arbitrary mixing of
unequal loss modes is assumed.

\section{Adaptive converse}\label{app:adaptive}

\subsection{Finite-replica bound}

\begin{lemma}[Heterogeneous adaptive bound]\label{lem:replica}
Fix a finite $n$-use protocol for \eqref{eq:product}, with finite
unconditional energies $x_{ij}$ and final retained state $\zeta_n$.
Then \eqref{eq:replicamain} holds for every positive integer $r$,
and \eqref{eq:adaptivegain} follows.
\end{lemma}

\begin{proof}
Represent the product channel in each round as $m$ consecutive
component calls, with no intervening processing. This is the same
channel because the component maps act on disjoint inputs. Before
subcall $(i,j)$, all unsent modes remain in Alice's memory and all
previous outputs are in Bob's memory. A trace-preserving operation on
other modes leaves the unconditional marginal of $A_{ij}$ unchanged.
Thus its energy is the original $x_{ij}$.

Denote the states immediately before and after this subcall by
$\rho^{(i,j)}$ and $\omega^{(i,j)}$, including every memory and stored
record. Run $r$ independent copies of the whole protocol, each with
its own outcomes and feedback. At the subcall in question, the
aggregate states are $(\rho^{(i,j)})^{\otimes r}$ and
$(\omega^{(i,j)})^{\otimes r}$. The $r$ transmitted modes experience
$\mathcal L_{\eta_j}^{\otimes r}$ and have total mean $rx_{ij}$.
By \cref{lem:importedblock},
\begin{align}
 \ER((\omega^{(i,j)})^{\otimes r})
 \le{}&\ER((\rho^{(i,j)})^{\otimes r})
       +r c_{\eta_j}(x_{ij})\nonumber\\
 &+g(r\eta_jx_{ij})+g(rq_jx_{ij}).
 \label{eq:subcall}
\end{align}
At each subcall only the transmitted systems change sides of the
Alice--Bob bipartition.

The intermediate REEs are finite: the initial state is separable,
and \eqref{eq:importedblock} with $s=1$ adds a finite amount at
each of the finitely many subcalls. Between-call operations are
fixed LOCC maps on the active quantum memories and classical
records, including their control by past outcomes. Their $r$-fold
products are again LOCC. Summing \eqref{eq:subcall}, using
monotonicity between calls and after the last call, gives
\eqref{eq:replicamain}.

For every finite $y\ge0$, $g(ry)/r\to0$. Divide the bound by $r$
and take $r\to\infty$, with $n$, $m$, and the protocol fixed, to
obtain \eqref{eq:adaptivegain}. The existence of the regularization
follows from finiteness and subadditivity. This limit precedes the
communication limit $n\to\infty$.
\end{proof}

Applying Jensen's inequality separately to each mode gives
\begin{equation}
 \frac1n\sum_{i,j}c_{\eta_j}(x_{ij})
 \le\sum_jc_{\eta_j}\!\left(\frac1n\sum_i x_{ij}\right)
 \le\Feta(N),
\end{equation}
which proves \eqref{eq:master}. Classical histories need not have
bounded entropy because the lemma uses only unconditional optical
means and keeps the histories as local records.

\subsection{Conversion into quantum and key bounds}

For completeness, we give the channel-independent testing argument.
A private state of key dimension $d$ is
\begin{equation}
 \begin{aligned}
 \gamma&=U_{\rm tw}(\Phi_d\otimes\theta_{S_AS_B})U_{\rm tw}^\dagger,\\
 \Pi_\gamma&=U_{\rm tw}(\Phi_d\otimes I_{S_AS_B})U_{\rm tw}^\dagger,
 \end{aligned}
 \label{eq:privatestate}
\end{equation}
where $U_{\rm tw}$ is a unitary controlled in the key basis
\cite{PrivateStates}. The privacy test obeys
$\Tr(\Pi_\gamma\sigma)\le d^{-1}$ for every state separable across
$K_AS_A:K_BS_B$ \cite{WildeTomamichelBerta}.
Trivial shields give the maximally entangled-state test.

\begin{lemma}[Regularized privacy-test bound]\label{lem:privacy}
If a state $\zeta$ has finite REE and
$\Tr(\Pi_\gamma\zeta)\ge1-\varepsilon$, $0\le\varepsilon<1$, then
\begin{equation}
 \ERi(\zeta)\ge(1-\varepsilon)\log_2d-h_2(\varepsilon).
 \label{eq:privacyree}
\end{equation}
\end{lemma}

\begin{proof}
Test $t$ copies independently and accept when at least
$k_t=\lfloor(1-\varepsilon-\alpha)t\rfloor$ tests pass, where
$0<\alpha<1-\varepsilon$. The acceptance $p_t$ on
$\zeta^{\otimes t}$ tends to one by the law of large numbers.
The product of any specified $k_t$ privacy tests is itself a
privacy test with key dimension $d^{k_t}$. A globally separable
comparison therefore passes that subset with probability at most
$d^{-k_t}$. Its probability of passing at least $k_t$ tests is at
most $\binom{t}{k_t}d^{-k_t}$, by the union bound. This remains a
valid upper bound when it exceeds one.

Binary data processing for relative entropy yields
\begin{equation}
 \ER(\zeta^{\otimes t})\ge
 p_t\left[k_t\log_2d-\log_2\binom{t}{k_t}\right]-h_2(p_t).
 \label{eq:thresholdtest}
\end{equation}
Divide by $t$, take $t\to\infty$, and then let $\alpha\downarrow0$.
The binomial entropy limit gives \eqref{eq:privacyree}.
\end{proof}

Combining \eqref{eq:privacyree} with \eqref{eq:master} proves
\eqref{eq:quantweak}. For operational key, retain local
purifications of the parties' instruments and postpone the final
key readout. Public messages remain copied to Eve, and subsequent
operations use the original active records. Tracing out the
additional local registers recovers the original protocol jointly
with Eve at every stage. Consequently every joint optical input
marginal, and hence every $x_{ij}$, is preserved. This is the
energy-preserving private-state reduction of
Ref.~\cite{Pirandola2026}, based on the private-state formalism
\cite{PrivateStates}.

With squared fidelity
$F(\rho,\sigma)=\norm{\sqrt\rho\sqrt\sigma}_1^2$, the
trace-distance criterion \eqref{eq:keyerror} implies fidelity at
least $(1-e)^2$ with an ideal classical key and the actual Eve
marginal. Uhlmann's theorem supplies a private-state comparison
for the retained output with the same fidelity lower bound
\cite{Uhlmann,Fuchs}. The associated privacy test therefore
accepts with probability at least $(1-e)^2$. Set
$\varepsilon=2e-e^2$ in \eqref{eq:privacyree} to obtain
\eqref{eq:keyweak}. The shields need no dimension or energy bound.

Taking vanishing error in \cref{eq:quantweak,eq:keyweak} proves
the converse part of \cref{thm:capacity}. Quantum teleportation
and the one-time pad transfer the bounds to $D_2$ and $P_2$.
At $N=0$, every transmitted input is vacuum and independent of
the memories, so all four capacities vanish.

\section{Finite-dimensional achievability}\label{app:achieve}

Fix an allocation $\boldsymbol N$ with $\sum_jN_j\le N$.
For a positive integer $s$, put $k_j=\lfloor sN_j\rfloor$.
Alice locally prepares $\bigotimes_j\Phi_{s,k_j}$ and transmits
the $s$ signal modes of pair $j$ through
$\mathcal L_{\eta_j}^{\otimes s}$. Arrange these signals in $s$
multimode uses. Their exact total photon number is
$\sum_jk_j\le sN$.

Bob measures the received total for each $j$ and announces it.
Conditional on losing $\ell_j$ photons, the resource has reverse
coherent information
\begin{equation}
 r_{s,k_j,\ell_j}=\log_2d_{s,k_j}-\log_2d_{s,\ell_j}\ge0.
 \label{eq:sectorhash}
\end{equation}
All conditional states are finite dimensional. Repeat the entire
block $t$ times. For each of the finitely many outcome classes,
keep a fixed number slightly below its expected frequency and
apply a finite-dimensional reverse-hashing code
\cite{DevetakWinter}. If a class is short, produce a separable
state in the same prescribed target dimension. The shortage
probability tends to zero, and the conditional hashing errors
also vanish. All $st$ multimode transmissions and all branches
are included in the rate.

At fixed $s$, this construction achieves arbitrarily close to
\begin{equation}
 \frac1s\sum_{j=1}^{m}
 \E_{L_j}\log_2\frac{d_{s,k_j}}{d_{s,L_j}},
 ~~ L_j\sim\Bin(k_j,q_j).
 \label{eq:sectorrate}
\end{equation}
The dimension asymptotic
$s^{-1}\log_2d_{s,\lfloor sx\rfloor}\to g(x)$ and
$L_j/s\to q_jN_j$ in probability imply that
\eqref{eq:sectorrate} tends to $\sum_jc_{\eta_j}(N_j)$.
For the expectation, the integrand divided by $s$ lies between
zero and $g(k_j/s)\le g(N_j)$; thus concentration and this
uniform bound justify the limit. When $N_j=0$ the corresponding
term is exactly zero.

The transmitted marginal $P_{s,k_j}/d_{s,k_j}$ is invariant
under permutations of its $s$ modes. At every multimode use,
the unconditional total input mean is
$\sum_jk_j/s\le N$. Vacuum padding handles other blocklengths.
Taking $t\to\infty$ first and then $s\to\infty$, and optimizing
over $\boldsymbol N$, proves the lower bound in \cref{thm:capacity}.
With positive weights $w_j$, the identical argument has per-use
mean cost $\sum_jw_jk_j/s\le\sum_jw_jN_j$ and proves the
weighted-energy achievability statement. Measuring the distilled
entanglement yields secret key with vanishing operational error;
teleportation and the one-time pad provide the remaining tasks.

\section{Allocation properties and asymptotics}\label{app:allocation}

Differentiating the thermal entropy yields
$g'(x)=\log_2(1+x^{-1})$ and
$g''(x)=-[\ln2\,x(x+1)]^{-1}$. Substitution gives
\cref{eq:slope,eq:curvature}. Strict negativity of the second
derivative, together with $\lim_{x\to\infty}c_\eta'(x)=0$,
implies $c_\eta'(x)>0$. At the origin,
\begin{equation}
 c_\eta'(x)=\eta\log_2(1/x)+q\log_2q+O(x),
 \label{eq:originslope}
\end{equation}
so the slope diverges for every $\eta>0$.

\begin{proof}[Proof of \cref{prop:allocation}]
Continuity on the compact feasible simplex gives existence, and
strict concavity gives uniqueness. Monotonicity exhausts the budget.
If a positive-transmissivity mode had zero allocation, transferring
a sufficiently small amount of energy from a populated mode would
gain an amount of order $t\log(1/t)$ while losing only $O(t)$.
Thus the optimizer is interior on $\sum_jN_j=N$.

The Lagrange multiplier equations are \eqref{eq:kkt}.
For every $\lambda>0$, the strictly decreasing derivative has a
unique inverse $x_j(\lambda)>0$. The sum
$\sum_jx_j(\lambda)$ decreases continuously from infinity to
zero as $\lambda$ increases from zero to infinity. It therefore
equals $N$ at one and only one multiplier.

For $z=(1-\eta)x>0$,
\begin{equation}
 \frac{\partial}{\partial\eta}c_\eta'(x)
 =\frac1{\ln2}\left[\ln(1+z^{-1})-\frac1{1+z}\right]>0.
 \label{eq:etamarginal}
\end{equation}
The last inequality follows by integrating $(1+t)^{-1}$ from
zero to $1/z$, which exceeds the endpoint lower rectangle.
Hence the larger-transmissivity mode must have a larger mean
to attain the same marginal value $\lambda$.

An interpolation of optimal feasible allocations at two budgets
is feasible at the interpolated budget. Concavity of the objective
then proves concavity of $\Feta$. Monotonicity is immediate.
The inverse derivative functions are smooth with nonzero
derivative, so the multiplier and optimizer are smooth for
$N>0$. Differentiating the value along the optimizer gives
$\Feta'(N)=\sum_j\lambda(N_j^*)'=\lambda$.
\end{proof}

\subsection{Low- and high-energy behavior}

Two asymptotic regimes give useful checks and design rules. Let
$\eta_{\max}=\max_j\eta_j$. At small total budget,
\begin{equation}
 \Feta(N)=\eta_{\max}N\log_2(1/N)+O(N),~~ N\downarrow0.
 \label{eq:small}
\end{equation}
At high energy, with the finite transmissivity list fixed,
\begin{align}
 \frac{N_j^*(N)}{N}
 &\longrightarrow
 \frac{\sqrt{\eta_j/q_j}}{\sum_k\sqrt{\eta_k/q_k}},
 \label{eq:highallocation}\\
 \Feta(N)
 &=\Beta-\frac{\left(\sum_j\sqrt{\eta_j/q_j}\right)^2}
                     {2\ln2\,N}+O(N^{-2}).
 \label{eq:highrate}
\end{align}
For the low-energy expansion,
\begin{equation}
 c_\eta(x)=\eta x\log_2(1/x)
       +\left(\frac{\eta}{\ln2}+q\log_2q\right)x+O(x^2).
 \label{eq:lowcomponent}
\end{equation}
Allocating all photons to a strongest mode gives the lower
bound in \eqref{eq:small}. Conversely, for $N<1$ and any
allocation with sum $N$,
$\sum_jN_j\log_2(1/N_j)\le N\log_2(m/N)$.
Using \eqref{eq:lowcomponent} and
$\eta_j\le\eta_{\max}$ proves the matching upper expansion.

At large $x$,
\begin{align}
 c_{\eta_j}(x)&=-\log_2q_j-\frac{a_j}{x}+O(x^{-2}),
 \label{eq:componentlarge}\\
 c_{\eta_j}'(x)&=\frac{a_j}{x^2}+O(x^{-3}).\label{eq:largeslope}
\end{align}
where $a_j=\eta_j/(2q_j\ln2)$.
As $N\to\infty$, the common multiplier tends to zero and all
optimal means diverge. Otherwise \eqref{eq:kkt} would keep
all means bounded. Inverting \eqref{eq:largeslope} gives
$N_j^*=\sqrt{a_j/\lambda}+O(1)$.
Summing yields
$\sqrt\lambda=(\sum_j\sqrt{a_j})/N+O(N^{-2})$ and hence
\begin{equation}
 N_j^*=\frac{\sqrt{a_j}}{\sum_k\sqrt{a_k}}\,N+O(1).
\end{equation}
Substituting in \eqref{eq:componentlarge} proves
\cref{eq:highallocation,eq:highrate}. The remainder constants
may depend on the fixed transmissivity list; the expansion is
not asserted uniformly as a transmissivity approaches an endpoint.

\section{Mean-energy strong-converse threshold}\label{app:meanstrong}

\subsection{Product-resource upper bound}

We state explicitly why the unconstrained converse extends to a
finite product. The bosonic simulation framework \cite{Simulation}
uses a finite-energy resource $\omega_j^\mu$ for each channel and
a separable comparison $\sigma_j^\mu$. We combine these resources
with the hypothesis-testing estimate stated below. If necessary, a
vanishing thermal regularization makes both states faithful
before applying the relative-entropy-variance bound. The
product channel is simulated with
\begin{equation}
 \Omega^\mu=\bigotimes_{j=1}^{m}\omega_j^\mu,~~
 \Sigma^\mu=\bigotimes_{j=1}^{m}\sigma_j^\mu.
\end{equation}
Across the aggregate Alice--Bob cut, $\Sigma^\mu$ is separable.
For these product states the relative entropy and its variance
satisfy
\begin{align}
 D(\Omega^\mu\Vert\Sigma^\mu)
    &=\sum_jD(\omega_j^\mu\Vert\sigma_j^\mu),\\
 V(\Omega^\mu\Vert\Sigma^\mu)
    &=\sum_jV(\omega_j^\mu\Vert\sigma_j^\mu).
 \label{eq:DV}
\end{align}
The second identity follows because the log-likelihood operator
is a sum of operators on independent tensor factors.
In the pure-loss limits of these resource comparisons,
\begin{equation}
 D(\Omega^\mu\Vert\Sigma^\mu)\longrightarrow\Beta,~~
 V(\Omega^\mu\Vert\Sigma^\mu)\longrightarrow0.
 \label{eq:DVlimit}
\end{equation}
The limits include removal of the faithful-state regularization.
The finiteness of $m$ permits the sums to pass through these limits.

For clarity, simulation is removed at fixed physical protocol.
Every component input of a protocol satisfying \eqref{eq:energy}
has mean at most $nN$. Energy-constrained convergence of each
component simulator, with an unrestricted reference
\cite{Simulation,Shirokov2018}, and a hybrid expansion give
\begin{equation}
 \norm{\zeta_n-\zeta_n^\mu}_1
 \le\sum_{i,j}\delta_{j,\mu}(nN)\longrightarrow0,
 \label{eq:simerror}
\end{equation}
where $\delta_{j,\mu}(E)$ is the supremum trace-norm error at
input mean at most $E$. The hybrid uses original prefixes and
simulated suffixes, so replacements are evaluated on actual
inputs. No energy bound on a simulated memory is needed.

The hypothesis-testing estimate of
Ref.~\cite{WildeTomamichelBerta}, applied to
$(\Omega^\mu)^{\otimes n}$ and $(\Sigma^\mu)^{\otimes n}$,
then gives the finite-error bound
\begin{equation}
 \begin{gathered}
 \log_2d\le n\Beta+C(\varepsilon),~~ 0<\varepsilon<1,\\
 C(\varepsilon)=\log_2 6+
        2\log_2\frac{1+\varepsilon}{1-\varepsilon},
 \end{gathered}
 \label{eq:unconstrainedbound}
\end{equation}
for an output with squared fidelity at least $1-\varepsilon$
to a private state of key dimension $d$. In taking the limits,
first remove simulation error at fixed $n$ and fixed protocol,
then remove the faithful-state regularization.
\Cref{eq:DV,eq:DVlimit} make the calculation identical to the
single-resource argument, with its entropy value replaced by
$\Beta$ and its limiting variance still zero.

Suppose a sequence has rate strictly above $\Beta$ and quantum
fidelity bounded below on a subsequence. Choose a fixed
$\varepsilon<1$ compatible with that lower bound.
Equation \eqref{eq:unconstrainedbound} contradicts the rate,
because $C(\varepsilon)/n\to0$. Thus quantum fidelity vanishes.
If the operational key error were bounded above by some
$e_0<1$ on a subsequence, the retained private-state fidelity
would be at least $(1-e_0)^2$, giving the same contradiction.
This proves the upper threshold for all four tasks.

\subsection{Matching threshold by vacuum randomization}

Fix $0<R<\Beta$ and $N>0$. Because $\Feta(E)\to\Beta$,
choose a finite $E>N$ with $R<\Feta(E)$.
The construction of Appendix~\ref{app:achieve} supplies an active
protocol sequence with budget $E$, fixed target dimension
$d_n$, rate approaching $R$, and quantum fidelity
$f_n^{\rm on}\to1$. It obeys the per-use unconditional budget.

Using a public flag, run the active protocol with probability
$p=N/E$, and otherwise send vacuum in all channel slots and
output a fixed product state $\ket{00}$ in the same target
registers. The unconditional total mean is at most $pE=N$
at each use. The resulting fidelity is
\begin{equation}
 f_n=p f_n^{\rm on}+(1-p)/d_n\longrightarrow p>0.
 \label{eq:dilution}
\end{equation}
Neither blocklength nor target dimension depends on the flag.
For key generation, the off branch outputs a fixed public key
value. Its distance from a uniform ideal key is $1-1/d_n$.
The public flag makes the branches orthogonal in Eve's record,
so the operational error tends to $1-p<1$.
Thus no threshold below $\Beta$ is possible.
Together with \eqref{eq:unconstrainedbound}, this proves
\cref{thm:meanstrong}.

\section{Spectral optimization and the continuum converse}\label{app:continuum}

\subsection{A finite list of adaptive component calls}

The continuum argument needs a version of the discrete converse before
any averaging over repeated uses. The following consequence of the
replica method makes that step explicit.

\begin{lemma}[Finite-list entanglement gain]\label{lem:finitelist}
Consider a finite ordered list of pure-loss calls with transmissivities
$\eta_1,\ldots,\eta_M$, arbitrary LOCC between calls, separable initial
state, and unconditional input means $x_1,\ldots,x_M<\infty$.
For the final retained state $\zeta$,
\begin{equation}
 \ERi(\zeta)\le\sum_{k=1}^{M}c_{\eta_k}(x_k).
 \label{eq:finitelist}
\end{equation}
The quantum and operational-key bounds are
\begin{align}
 (1-\varepsilon)\log_2d
 &\le\sum_kc_{\eta_k}(x_k)+h_2(\varepsilon),\label{eq:listquant}\\
 (1-e)^2\log_2d
 &\le\sum_kc_{\eta_k}(x_k)+h_2(2e-e^2).\label{eq:listkey}
\end{align}
\end{lemma}

\begin{proof}
Run $r$ independent replicas of the complete list protocol. At slot $k$,
the $r$ signals have the same transmissivity and aggregate mean $rx_k$.
The block lemma and LOCC monotonicity give
\begin{equation}
 \begin{aligned}
 \ER(\zeta^{\otimes r})\le{}&r\sum_kc_{\eta_k}(x_k)\\
 &+\sum_k[g(r\eta_kx_k)+g(r(1-\eta_k)x_k)].
 \end{aligned}
 \label{eq:listreplica}
\end{equation}
Divide by $r$ and let $r\to\infty$, with the entire list fixed.
The last sum is $o(r)$ because there are finitely many finite-energy
terms. The endpoint extensions are in Appendix~\ref{app:endpoints}.
The testing argument of \cref{lem:privacy} and its energy-preserving
key reduction prove \cref{eq:listquant,eq:listkey}.
\end{proof}
The bound allows arbitrary input correlations and feedback. In
particular, setting $M=M(T)$ causes no problem: $r\to\infty$ is
taken at each fixed $T$, before the limit $T\to\infty$. There is
no residual replica term to control uniformly in $M(T)$.

\subsection{Scalar dual and existence of the optimizer}

For $s>0$, introduce the nonnegative scalar dual function
\begin{equation}
 \phi_\eta(s)=\sup_{x\ge0}[c_\eta(x)-sx].
 \label{eq:phidef}
\end{equation}
For $\eta>0$, strict concavity and the endpoint slopes give a unique
maximizer $x_\eta(s)>0$ characterized by $c_\eta'(x_\eta(s))=s$.
Set $x_0(s)=0$ and $\phi_0(s)=0$. Since $0\le c_\eta\le g$,
\begin{equation}
 \begin{gathered}
 0\le\phi_\eta(s)\le\phi_1(s)
       =-\log_2(1-2^{-s}),\\
 0\le x_\eta(s)\le x_1(s)=\frac1{2^s-1}.
 \end{gathered}
 \label{eq:dualenvelope}
\end{equation}
The second inequality follows from $c_\eta'(x)\le g'(x)$.
These functions are continuous in $(\eta,s)$ for
$0\le\eta\le1$ and $s>0$; at $\eta=0$ the optimizing occupation
tends to zero. The maximization can be restricted to the compact
interval $[0,x_1(s)]$, which also proves continuity of $\phi$.

For every feasible profile and $\lambda>0$,
\begin{equation}
 \begin{aligned}
 \int_{\mathcal B}\frac{\mathrm d\omega}{2\pi}
 c_{\eta(\omega)}(\nu(\omega))
 \le{}&\lambda P\\
 &+\int_{\mathcal B}\frac{\mathrm d\omega}{2\pi}
       \phi_{\eta(\omega)}(\lambda\hbar\omega).
 \end{aligned}
 \label{eq:dualupper}
\end{equation}
The integral on the right is finite even on $(0,\infty)$ because
$\phi_1(\lambda\hbar\omega)$ is integrable at both endpoints.
Define
\begin{equation}
 \mathcal P(\lambda)=
 \int_{\mathcal B}\frac{\mathrm d\omega}{2\pi}
       \hbar\omega x_{\eta(\omega)}(\lambda\hbar\omega).
 \label{eq:Plambda}
\end{equation}
The occupation bound in \eqref{eq:dualenvelope} makes this finite.
If the transmitting set has positive measure, $\mathcal P$ is
continuous and strictly decreasing. Continuity follows by dominated
convergence on any compact interval of positive multipliers, and strict
decrease holds on the transmitting set. The same domination gives
$\mathcal P(\lambda)\to0$ as $\lambda\to\infty$.
For $\lambda\downarrow0$, the occupation diverges at every transmitting
frequency; monotone convergence on a bounded positive-frequency
subset of positive measure gives $\mathcal P(\lambda)\to\infty$.
Thus exactly one $\lambda$ satisfies $\mathcal P(\lambda)=P$.

At this multiplier, $\nu_*(\omega)=x_{\eta(\omega)}(\lambda\hbar\omega)$
is feasible and saturates the pointwise inequality defining $\phi$.
It attains \eqref{eq:dualupper}, proving the exact dual identity
\begin{equation}
 \Cspec(P)=\inf_{\lambda>0}\left\{
 \lambda P+
 \int_{\mathcal B}\frac{\mathrm d\omega}{2\pi}
       \phi_{\eta(\omega)}(\lambda\hbar\omega)\right\}.
 \label{eq:continuousdual}
\end{equation}
Strict concavity gives uniqueness almost everywhere on the transmitting
set. The budget is exhausted there, so no positive power can be assigned
to the completely lost set by an optimizer. Concavity of the value
follows either from feasible interpolation or from the dual formula.
The unique minimizing multiplier is its derivative with respect to $P$.
Since this multiplier is positive, the capacity is increasing for
$P>0$. The capacity is zero at $P=0$ or when $\eta=0$ almost
everywhere.

For completeness,
\begin{equation}
 \int_0^\infty\frac{\mathrm d\omega}{2\pi}
 \phi_1(\lambda\hbar\omega)
 =\frac{\pi}{12\lambda\hbar(\ln2)^2}.
 \label{eq:idealdual}
\end{equation}
Indeed, expand $-\ln(1-e^{-t})$ in its positive power series and
integrate termwise, obtaining $\sum_{r\ge1}r^{-2}=\pi^2/6$.
Minimizing $\lambda P$ plus the right side of \eqref{eq:idealdual}
proves \eqref{eq:idealrate} and also continuity of the capacity at $P=0$.

\subsection{Lattice convergence, including both spectral tails}

Fix $\lambda>0$, write $\Delta_T=2\pi/T$, and put
\begin{equation}
 f_\lambda(\omega)=\phi_{\eta(\omega)}(\lambda\hbar\omega),
 ~~ G_\lambda(\omega)=\phi_1(\lambda\hbar\omega).
 \label{eq:spectralenvelope}
\end{equation}
On a compact positive-frequency interval, $f_\lambda$ is bounded and
piecewise continuous, so its lattice sums converge to its integral.
The function $G_\lambda$ is nonnegative, decreasing, and integrable.
For the full positive spectrum, right-endpoint comparison gives
\begin{align}
 \frac1T\sum_{k:\,k\Delta_T\le a}f_\lambda(k\Delta_T)
 &\le\int_0^a\frac{\mathrm d\omega}{2\pi}G_\lambda(\omega),
 \label{eq:IRtail}\\
 \frac1T\sum_{k:\,k\Delta_T\ge b}f_\lambda(k\Delta_T)
 &\le\int_{b-\Delta_T}^{\infty}\frac{\mathrm d\omega}{2\pi}
                 G_\lambda(\omega),
 \label{eq:UVtail}
\end{align}
where $b>\Delta_T$ in the second line. First make $a$ small and $b$
large, and then take $T\to\infty$. These bounds control both tails
independently of the chosen input profile. Combining them with compact
Riemann convergence proves
\begin{equation}
 \frac1T\sum_{k:\,\omega_k\in\mathcal B}
      \phi_{\eta(\omega_k)}(\lambda\hbar\omega_k)
 \longrightarrow
 \int_{\mathcal B}\frac{\mathrm d\omega}{2\pi}
      \phi_{\eta(\omega)}(\lambda\hbar\omega).
 \label{eq:dualRiemann}
\end{equation}
Finite band endpoints do not affect the limit. In particular, no
assumption of uniform convergence of the optimizing occupations is
required near zero frequency or in the ultraviolet tail.

Apply \cref{lem:finitelist} to any finite duration-$T$ protocol.
Its actual unconditional means obey \eqref{eq:powerfinite}, and hence
\begin{equation}
 \begin{aligned}
 \sum_kc_{\eta(\omega_k)}(N_{T,k})
 \le{}&\lambda PT\\
 &+\sum_{k:\,\omega_k\in\mathcal B}
       \phi_{\eta(\omega_k)}(\lambda\hbar\omega_k).
 \end{aligned}
 \label{eq:latticedual}
\end{equation}
The sum on the right includes unselected modes, which can only increase
it because $\phi\ge0$. It converges for every fixed $T$ by the same
envelope. Substitute \eqref{eq:latticedual} into
\cref{eq:listquant,eq:listkey}, divide by $T$, and take vanishing error
and $T\to\infty$. Equation \eqref{eq:dualRiemann} gives an upper bound
equal to the dual objective at every $\lambda>0$. Taking the infimum
and using \eqref{eq:continuousdual} proves the continuum converse.
All arguments apply to finite target registers. The same bounds hold
for trace-norm limits of finite protocols that preserve the power
budget and these registers.

\section{Achievability by spectral interval coding}\label{app:spectralcoding}

We prove that the variational rate is attainable without preshared
entanglement. Fix $P>0$ and a target rate strictly below $\Cspec(P)$.
By continuity in power, choose $P'<P$ whose capacity still exceeds the
target. Let $\nu_*$ be its optimizing profile. Since
$0\le c_\eta(\nu_*)\le g(\nu_*)$ and the Gibbs bound
\eqref{eq:dualupper} is integrable, the capacity integral can be
approximated from below on a compact subband
$K\subset\mathcal B\cap[a,b]$, with $0<a<b<\infty$.
For a finite union of available bands, take the corresponding finite
union of compact pieces. The contribution of omitted intervals can be
made arbitrarily small.

Partition $K$ into finitely many intervals $I_\ell$ of lengths
$|I_\ell|$. Let $\bar\omega_\ell$ be the upper endpoint and let
$\eta_\ell$ be a lower bound on $\eta(\omega)$ throughout the interval.
Choose it as the infimum. On intervals away from discontinuities this
converges to the local value as the mesh decreases; intervals meeting
the finitely many discontinuities can be omitted with vanishing total
length. Define the constant occupation
\begin{equation}
 \nu_\ell=x_{\eta_\ell}(\lambda'\hbar\bar\omega_\ell),
 \label{eq:binoccupation}
\end{equation}
where $\lambda'$ is the multiplier for $P'$.
On $[a,b]$, the relevant occupations and dual parameters are bounded,
and the continuity established in Appendix~\ref{app:continuum} yields
\begin{align}
 \sum_\ell\frac{|I_\ell|}{2\pi}
       \hbar\bar\omega_\ell\nu_\ell
 &\longrightarrow
 \int_K\frac{\mathrm d\omega}{2\pi}\hbar\omega\nu_*(\omega),
 \label{eq:binpower}\\
 \sum_\ell\frac{|I_\ell|}{2\pi}c_{\eta_\ell}(\nu_\ell)
 &\longrightarrow
 \int_K\frac{\mathrm d\omega}{2\pi}
          c_{\eta(\omega)}(\nu_*(\omega)).
 \label{eq:binrate}
\end{align}
Select a sufficiently fine finite partition so the first sum is strictly
below $P$ and the second remains above the target rate with a margin.

The number $M_{T,\ell}$ of lattice modes in $I_\ell$ satisfies
\begin{equation}
 M_{T,\ell}=\frac{T|I_\ell|}{2\pi}+O(1).
 \label{eq:bincount}
\end{equation}
If $\eta_\ell>0$, Bob applies a local pure-loss map of transmissivity
$\eta_\ell/\eta(\omega_k)$ to each output in the interval. Its
composition with the physical channel is exactly
$\mathcal L_{\eta_\ell}$. Thus all modes in this interval implement
the same effective channel. Intervals with $\eta_\ell=0$ need not
be used. Giving all discarded attenuator outputs to Eve, if necessary,
only enlarges the adversary's information; the resulting channel still
has a vacuum pure-loss dilation at the effective transmissivity.

Choose a fixed positive integer sector length $s_\ell$ and put
$k_\ell=\lfloor s_\ell\nu_\ell\rfloor$.
Use $t_{T,\ell}=\lfloor M_{T,\ell}/s_\ell\rfloor$ disjoint groups
of $s_\ell$ frequencies to prepare and send the sector states of
Appendix~\ref{app:achieve}. No lattice mode is used twice. Every signal
marginal in a group has mean $k_\ell/s_\ell\le\nu_\ell$.
Because its physical frequency is at most $\bar\omega_\ell$, the
expected energy spent in the interval is at most
$M_{T,\ell}\hbar\bar\omega_\ell\nu_\ell$.
Equations \eqref{eq:binpower} and \eqref{eq:bincount}, with the strict
power margin, imply the budget $PT$ for all sufficiently large $T$.
The unused frequencies carry vacuum.

At fixed sector length, $t_{T,\ell}\to\infty$, so the finite-dimensional
reverse-hashing construction produces a fixed target size with vanishing
error. Its asymptotic rate per used frequency is arbitrarily close to
\begin{equation}
 \frac1{s_\ell}\E_{L_\ell}
       \log_2\frac{d_{s_\ell,k_\ell}}{d_{s_\ell,L_\ell}},
 ~~ L_\ell\sim\Bin(k_\ell,1-\eta_\ell).
 \label{eq:binhash}
\end{equation}
For sufficiently large but fixed $s_\ell$, this differs from
$c_{\eta_\ell}(\nu_\ell)$ by an arbitrarily small amount. There are
only finitely many intervals, so their errors sum to zero and their
fixed-dimensional targets can be combined. Equation \eqref{eq:bincount}
therefore gives the total rate in \eqref{eq:binrate}.

The choices occur in this order: select a strict power margin, a compact
subband, a finite partition, and finite sector lengths; then let
$T\to\infty$ to perform hashing. These choices attain every rate below
$\Cspec(P)$. A diagonal sequence removes the margins if desired.
The construction first distributes entanglement. Measuring it,
teleporting, or using the one-time pad supplies the other three tasks
without additional optical energy. This completes
\cref{thm:continuum} for all four operational rates.

\section{Evaluation of the broadband coefficient}\label{app:broadband}

Put $q=1-\eta$, initially with $0<\eta<1$, and use natural-logarithm
entropy
\begin{equation}
 \begin{split}
 G(u)&=(u+1)\ln(u+1)-u\ln u,\\
 h_\eta(u)&=G(u)-G(qu).
 \end{split}
 \label{eq:naturalentropy}
\end{equation}
Then $h_\eta=(\ln2)c_\eta$ and
\begin{equation}
 h_\eta'(u)=\ln(1+u^{-1})-q\ln(1+(qu)^{-1}).
 \label{eq:naturalslope}
\end{equation}
This derivative decreases continuously from infinity to zero, and
$v_\eta$ in \eqref{eq:veta} is its inverse.
Writing $\Omega=(\lambda\hbar\ln2)^{-1}$ in
\eqref{eq:continuouskkt} gives $\nu_*(\omega)=v_\eta(\omega/\Omega)$.

The endpoint estimates are
\begin{align}
 v_\eta(x)&\sim\sqrt{\frac{\eta}{2qx}},&&x\downarrow0,
 \label{eq:vinfrared}\\
 v_\eta(x)&\sim q^{q/\eta}e^{-x/\eta},&&x\to\infty.
 \label{eq:vultraviolet}
\end{align}
They follow by expanding \eqref{eq:naturalslope} at large and small
occupation, respectively. In particular, both the power and rate
integrals converge. Let
\begin{equation}
 A_\eta=\int_0^\infty x v_\eta(x)\,\mathrm dx.
 \label{eq:Adef}
\end{equation}
Changing variables from $x$ to $u=v_\eta(x)$ and integrating by parts
gives
\begin{equation}
 A_\eta=\frac12\int_0^\infty[h_\eta'(u)]^2\,\mathrm du.
 \label{eq:Aparts}
\end{equation}
The boundary terms $u[h_\eta'(u)]^2$ vanish at zero and infinity.
A second integration by parts gives
\begin{equation}
 \int_0^\infty h_\eta(v_\eta(x))\,\mathrm dx=2A_\eta.
 \label{eq:rateparts}
\end{equation}
Here $\mathrm d[h_\eta(v_\eta(x))]/\mathrm dx=xv_\eta'(x)$,
and \cref{eq:vinfrared,eq:vultraviolet} justify both integrations.
Consequently
\begin{equation}
 P=\frac{\hbar\Omega^2}{2\pi}A_\eta,
 ~~
 \mathcal C_\eta^{\mathrm{bb}}(P)
     =\frac{\Omega A_\eta}{\pi\ln2}.
 \label{eq:powerandrate}
\end{equation}
Eliminating $\Omega$ proves \eqref{eq:broadband}, once $A_\eta$
has been evaluated.

For $a,b>0$, consider the elementary convergent integral
\begin{equation}
 I(a,b)=\int_0^\infty
       \ln(1+a/u)\ln(1+b/u)\,\mathrm du.
 \label{eq:Iab}
\end{equation}
Differentiating with respect to $a$ and then $b$ yields
\begin{equation}
 \frac{\partial^2 I}{\partial a\partial b}
 =\int_0^\infty\frac{\mathrm du}{(u+a)(u+b)}
 =\frac{\ln(b/a)}{b-a},
 \label{eq:Iderivative}
\end{equation}
with the continuous value $1/a$ at $a=b$.
Integrating in $b$ from zero gives
$I_a=\pi^2/6-\Li_2(1-b/a)$. Homogeneity,
$I(ta,tb)=tI(a,b)$, then gives
\begin{equation}
 \begin{aligned}
 I(a,b)={}&a\left[\frac{\pi^2}{6}-\Li_2(1-b/a)\right]\\
 &+b\left[\frac{\pi^2}{6}-\Li_2(1-a/b)\right].
 \end{aligned}
 \label{eq:Iclosed}
\end{equation}
All dilogarithms here have real arguments at most one.
Expanding the square in \eqref{eq:Aparts} gives
\begin{align}
 A_\eta
 &=\tfrac12[I(1,1)+q^2I(q^{-1},q^{-1})-2qI(1,q^{-1})]
 \nonumber\\
 &=\Li_2(\eta)+q\Li_2(-\eta/q).
 \label{eq:Aintermediate}
\end{align}
The real dilogarithm identity \cite{DLMF}
\begin{equation}
 \Li_2(-\eta/q)=-\Li_2(\eta)-\tfrac12\ln^2q
 \label{eq:dilogidentity}
\end{equation}
now proves \eqref{eq:Aeta}.

At $\eta=1$, $v_1(x)=(e^x-1)^{-1}$ and
$A_1=\int_0^\infty x/(e^x-1)\,\mathrm dx=\pi^2/6$.
At $\eta=0$ the channel has zero capacity, and the closed expression
has the continuous value zero. Expansion at small $\eta$ gives
\begin{equation}
 A_\eta=\tfrac12\eta^2+\tfrac14\eta^3+O(\eta^4).
 \label{eq:Asmall}
\end{equation}
Also $A_\eta'=\Li_2(\eta)+\tfrac12\ln^2(1-\eta)>0$,
which proves the monotonicity and asymptote stated for $\kappa$.
For $\eta=1/2$,
$A_{1/2}=\pi^2/24-\tfrac12\ln^22$,
yielding a numerical check in \eqref{eq:kappachecks}.

\section{Continuum mean-power strong converse}\label{app:continuousstrong}

The unconstrained testing argument in Appendix~\ref{app:meanstrong} also applies
to a finite ordered list of distinct component channels: use one
resource per component rather than $n$ identical product resources.
For a quantum or retained private-state output with squared target
fidelity at least $1-\varepsilon$, $0<\varepsilon<1$, it gives
\begin{equation}
 \log_2d_T\le\sum_{k\in S_T}b(\eta(\omega_k))+C(\varepsilon),
 \label{eq:spectralstrongfinite}
\end{equation}
where $S_T$ is the finite selected set and $C(\varepsilon)$ is the same
dimension-independent function as in \eqref{eq:unconstrainedbound}.
The relative entropy is the sum of the component values and the
limiting variance is zero. Simulation is removed for each fixed finite
protocol before $T$ increases. Thus \eqref{eq:spectralstrongfinite}
has no remainder depending on the number of selected modes.

When \eqref{eq:spectraltrace} holds with finite $\Bspec$, enlarge the
nonnegative sum in \eqref{eq:spectralstrongfinite} to all available
lattice modes. If a rate above $\Bspec$ had quantum fidelity bounded
below on a subsequence, a fixed $\varepsilon<1$ would give a
contradiction after division by $T$. For operational key, an error
bounded above by $e_0<1$ gives retained private-state fidelity at least
$(1-e_0)^2$ and the same contradiction. This proves the upper threshold.
For compact bands the trace condition is ordinary Riemann convergence
when $\eta$ stays below one. A decreasing integrable envelope on
$(0,\infty)$ gives the same two tail bounds as
\cref{eq:IRtail,eq:UVtail}, now for $b(\eta(\omega))$.

For the matching lower threshold, first note that
\begin{equation}
 \lim_{E\to\infty}\Cspec(E)=\Bspec,
 \label{eq:highpowerlimit}
\end{equation}
including the value $+\infty$.
The upper bound is pointwise $c_\eta(x)\le b(\eta)$.
For the lower bound, choose a bounded positive-frequency subband $K$
and set its occupation to a constant $L$, with zero occupation outside.
Its power is finite for every finite $L$. Monotone convergence as
$L\to\infty$ gives the integral of $b(\eta)$ on $K$; exhausting
$\mathcal B$ by such sets proves \eqref{eq:highpowerlimit}.

Fix $0<R<\Bspec$ and $P>0$. Choose a finite $E>P$ with
$R<\Cspec(E)$. The continuum achievability theorem supplies an active
code sequence of power at most $E$ and rate tending to $R$, with
vanishing error and a fixed target dimension $d_T$ at each $T$.
Run it with public probability $p=P/E$; on the other branch send vacuum
in the same slots and output a fixed product state or fixed public key
in the same target registers. The unconditional power is at most $P$.
Its quantum fidelity tends to $p$, and its operational key error tends
to $1-p$. Thus no rate below $\Bspec$ is a strong-converse threshold.
If $\Bspec=+\infty$, this argument applies to every finite $R$ and
already proves the infinite threshold, without invoking a spectral
trace condition. At $P=0$, all optical inputs are vacuum and both the
capacity and threshold are zero.

\subsection{Example: exponentially decaying transmission}

For example, a decaying profile
$\eta(\omega)=\eta_0e^{-\omega/\omega_c}$, with $0<\eta_0<1$,
has an integrable decreasing unconstrained-rate density and therefore
\begin{equation}
 \mathcal C_2^\dagger(P)=
 \frac{\omega_c}{2\pi\ln2}\Li_2(\eta_0),~~ P>0.
 \label{eq:exponentialthreshold}
\end{equation}
Expanding $-\ln(1-\eta_0e^{-\omega/\omega_c})$ and integrating each
positive term gives this expression. Its finite-power capacity is
determined by \cref{eq:continuouscapacity,eq:continuouskkt}.

\section{Endpoint modes and normalization checks}\label{app:endpoints}

A completely lost mode, $\eta_j=0$, is a vacuum
replacement channel and has $c_0(x)=0$. It can be omitted
from an optimal allocation whenever a transmitting mode is present.
If every mode is completely lost, all capacities and
thresholds are zero.

A perfectly transmitting mode, $\eta_j=1$, has
$c_1(x)=g(x)$. The block gain \eqref{eq:importedblock}
extends directly to this endpoint. On a fixed photon
sector, ordinary finite-dimensional teleportation uses
a resource of entanglement $\log_2d_{s,k}$.
For a random sector number of mean $sx$, its average
cost is at most $s g(x)$: the entropy of the
sector-uniform state is $H(K)+\E\log_2d_{s,K}$
and is at most the maximum entropy $s g(x)$ at that
total mean. Output number coherence costs at most
$g(sx)$. Thus \eqref{eq:importedblock} holds with
$\eta=1$ and $g(s(1-\eta)x)=0$.
The replica proof and sector-hashing lower bound
then give \eqref{eq:capacity} with $c_1=g$.

The allocation remains strictly concave on the
positive-transmissivity modes. Every such mode receives
positive energy for $N>0$; the derivative in
\eqref{eq:kkt} for an ideal mode is simply $g'(N_j)$.
The high-energy expansion \eqref{eq:highrate} is
restricted to strictly lossy modes. If any ideal mode
is present, $\Feta(N)\to\infty$ as $N\to\infty$,
and vacuum randomization gives
$C_2^\dagger(\Leta,N)=+\infty$ for every $N>0$.

At zero total budget all transmitted inputs are
vacuum, even when ideal modes are present; all four
capacities and thresholds are zero. For $m=1$ the
formulas reduce to the single-mode results. For
identical transmissivities, \eqref{eq:identical}
also follows by viewing $n$ multimode uses as
$nm$ scalar uses with average budget $N/m$.
For strictly lossy modes, sending $N\to\infty$
in \eqref{eq:F} recovers $\Beta$.

\section{Structure of the fading proofs}
\label{fd:app:map}
There are two separate converse mechanisms. The observed models
use the pure-loss sector bound inside replicas of the complete
adaptive protocol. The unobserved model uses an exact loss-count
extension and a number-pinching chain rule. The common operational
link is the relative entropy of entanglement and the target-test
bound in Appendix~\ref{fd:app:tools}.

\begin{figure*}[t]
\centering
\includegraphics[width=0.94\textwidth]{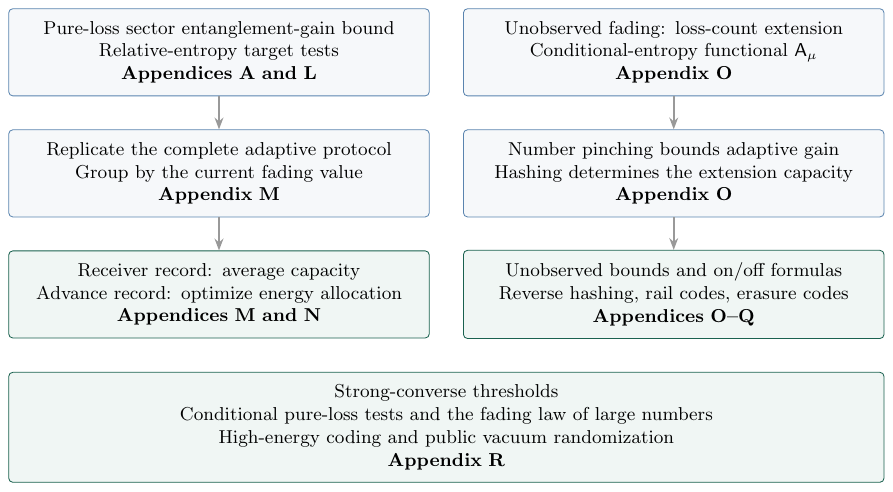}
\caption{Structure of the fading proofs, with the relevant appendices
listed in each box. The observed models use the
pure-loss replica estimate and optimize according to the timing of the
record. The unobserved model uses a loss-count extension and a pinching
bound, together with reverse-hashing and rail-code lower bounds.
The two branches share the operational target tests. A separate
argument conditions on the fading sequence to prove upper
strong-converse thresholds; high-energy coding and public
randomization give the matching thresholds in the stated cases.}
\label{fd:fig:proof}
\end{figure*}
The two chains are logically independent apart from their use of
the entanglement measure and the fact that a public fading record
provides an upper bound on the unobserved model. The strong-converse
argument is developed separately in Appendix~\ref{fd:app:strong}.

\section{Entanglement tools and operational converses}
\label{fd:app:tools}
We retain the relative-entropy conventions of Appendix~\ref{app:structure}
and record the additional flag identity used for fading. Thus
\begin{equation}
\ER(A:B)_\rho=\inf_{\sigma\in\mathrm{SEP}(A:B)}D(\rho\|\sigma),
\label{fd:eq:ree}
\end{equation}
where separability means the trace-norm closed convex hull of product
states \cite{Vedral1997,VedralPlenio}.
Its regularization is
\(\ERi(\rho)=\lim_{r\to\infty}r^{-1}\ER(\rho^{\otimes r})\).
The limit exists by subadditivity. Both quantities are monotone under
LOCC. A finite classical label copied to both laboratories obeys the
unregularized identity
\begin{equation}
\ER\!\left(\sum_jp_j\proj j_{F_A}\otimes\proj j_{F_B}
\otimes\rho_j\right)=\sum_jp_j\ER(\rho_j).
\label{fd:eq:flags}
\end{equation}
For the upper bound, use block-diagonal separable comparison states.
For the lower bound, dephase the comparison state in the local flags
and apply the direct-sum identity for relative entropy. Mismatched
flag blocks cannot improve the infimum. We will not require an
analogous identity for \(\ERi\).

\subsection{Targets and regularization}
A maximally entangled target of Schmidt rank \(d\), or a private
state of key dimension \(d\), admits a test \(\Pi\) which accepts
the target with certainty and satisfies
\(\Tr(\Pi\sigma)\leq d^{-1}\) for every separable \(\sigma\).
For private states, use the privacy test described in
Appendix~\ref{app:adaptive}. Products of these tests have the same
property with the product key dimension.

Suppose the actual retained state \(\zeta\) passes this test with
probability \(t\). Test each copy of \(\zeta^{\otimes r}\)
separately and let \(z\in\{0,1\}^r\) record the outcomes.
For an arbitrary separable comparison state across the two full
laboratories, the outcome probability \(q_z\) satisfies
\begin{equation}
q_z\leq d^{-|z|}.
\end{equation}
Indeed, discard the failure tests and apply the product privacy-test
bound on the \(|z|\) successful positions. The true outcome law
is a product of Bernoulli variables of mean \(t\). Data processing
then gives
\begin{align}
D(\zeta^{\otimes r}\|\sigma)
&\geq D(p_z\|q_z)\nonumber\\
&\geq rt\log_2d-rh_2(t).
\end{align}
Taking the separable infimum and regularizing yields
\begin{equation}
\ERi(\zeta)\geq t\log_2d-h_2(t).
\label{fd:eq:target}
\end{equation}
The same estimate with \(r=1\) bounds \(\ER\).
For a sequence of reliable quantum or private-state targets,
\(t\to1\), so a bound \(\ERi(\zeta_n)\leq nC\), or
\(\ER(\zeta_n)\leq nC\), proves the corresponding weak
capacity converse.

The standard key-to-private-state reduction retains local
purifications of the parties' processing and measurements
\cite{Horodecki2009,PrivateStates}. Public records remain public and are
copied to Eve. Tracing out the extra local registers recovers the
original protocol at every step, including each optical input.
Consequently the reduction preserves the unconditional energies in
Eq.~\eqref{fd:eq:budget}. Vanishing operational key error gives a
private-state test with \(t\to1\); no bound on the shield
dimension is required.

\subsection{The pure-loss input to the observed converses}
For the observed fading converses, we use the block estimate of
\cref{lem:importedblock}, restated here with the endpoint
extensions of Appendix~\ref{app:endpoints} \cite{Pirandola2026}. If \(s\) modes of
total mean photon number \(sx\) are transmitted through
\(\mathcal L_\eta^{\otimes s}\), with arbitrary retained memories
\(X,Y\), then
\begin{align}
\ER(X:B^sY)_\omega
\leq{}&\ER(XA^s:Y)_\rho+s c_\eta(x)\nonumber\\
&+g(s\eta x)+g(s(1-\eta)x).
\label{fd:eq:blockgain}
\end{align}
The two scalar terms are sublinear in \(s\) for fixed \(x\).
The zero-transmission endpoint has no entanglement gain; the
identity endpoint follows from the same sector simulation with
zero loss count. Thus the bound is valid on the closed interval
\(0\leq\eta\leq1\). We only apply it to states of finite input
REE, which holds inductively from a separable initial preparation.
The distinction between this memory-inclusive bound and a
capacity formula for a single attenuator is essential.

\section{Proof of the observed-fading capacity formulas}
\label{fd:app:observed}
\subsection{Finite alphabets: adaptive converse}
First let \(\mu=\sum_{j=1}^d w_j\delta_{\eta_j}\), omitting
zero weights. Fix an entire \(n\)-use protocol, and run \(r\)
independent replicas. All adaptive operations and all histories
inside a replica retain their original order. At round \(i\),
condition on the vector of current fading labels in the replicas.
Let \(r_j\) be the number of replicas with label \(j\).

For receiver information, the current label is independent of the
already prepared input and the whole previous history. Every
transmitted marginal in group \(j\) therefore has mean \(x_i\).
For advance information, let \(x_{ij}\) be the input mean after
the protocol's local processing conditioned on current label
\(j\), averaging over all past histories. Independence of the
replicas makes the total mean in this group \(r_jx_{ij}\).
We use \(x_{ij}=x_i\) in the receiver case.

Apply Eq.~\eqref{fd:eq:blockgain} to each group in this single
round, treating all other systems as retained memories.
Conditioned on the current label vector, the entanglement gain
is at most
\begin{equation}
\sum_{j:r_j>0}\left[
r_jc_{\eta_j}(x_{ij})+
g(r_j\eta_jx_{ij})+g(r_j(1-\eta_j)x_{ij})\right].
\end{equation}
Use Eq.~\eqref{fd:eq:flags} before and after the transmissions
to average this inequality. In the advance-information model,
the label-controlled preparation is LOCC and does not increase
the average input REE. Since \(\E r_j=rw_j\) and
\(r_j\leq r\), summing over rounds gives
\begin{align}
\ER(\zeta_n^{\otimes r})
\leq{}&r\sum_{i,j}w_jc_{\eta_j}(x_{ij})
+2\sum_{i,j}g(rx_{ij}).
\label{fd:eq:replicafading}
\end{align}
Each \(x_{ij}\) is finite: the weights are positive and the
unconditional energy is finite. Take \(r\to\infty\), with the
protocol and \(n\) fixed, to obtain
\begin{equation}
\ERi(\zeta_n)\leq\sum_{i,j}w_jc_{\eta_j}(x_{ij}).
\label{fd:eq:regularizedfading}
\end{equation}
Grouping occurs only across independent replicas at the same round
and preserves the chronological order of every adaptive run.

The function \(c_\eta\) is increasing and concave. In the
receiver model Jensen's inequality bounds the right-hand side
of Eq.~\eqref{fd:eq:regularizedfading} by
\(n\sum_jw_jc_{\eta_j}(N)\). In the advance model set
\(\bar x_j=n^{-1}\sum_i x_{ij}\). The same inequality bounds
it by \(n\sum_jw_jc_{\eta_j}(\bar x_j)\), with
\(\sum_jw_j\bar x_j\leq N\). Equation~\eqref{fd:eq:target}
therefore proves the two converses for a finite alphabet.

\subsection{Finite alphabets: achievability}
In the receiver model Alice repeatedly transmits half of a
two-mode squeezed vacuum of local mean \(N\). Conditioned on
label \(j\), the reverse coherent information is
\(g(N)-g((1-\eta_j)N)\). The parties sort their outputs by
the public labels and apply reverse hashing in each class
\cite{Reverse,DirectReverse,DevetakWinter}.
The asymptotic class frequencies are \(w_j\). Choosing a
fixed target rate strictly below their weighted sum makes the
probability of an insufficient class size tend to zero.
Such events may be declared failures; their slots and energies
are still included in the protocol's accounting.

For advance information, choose an admissible allocation
\(x_j\), and use a squeezed input of mean \(x_j\) whenever
the current label is \(j\). The same conditional distillation
achieves \(\sum_jw_jc_{\eta_j}(x_j)\). Truncated Schmidt
distributions and finite-dimensional hashing approach these
rates while respecting the energy bound; entropy convergence is
controlled by energy-constrained continuity bounds \cite{Winter2016}.
Optimization proves
the two finite-alphabet identities.

\subsection{General Borel fading laws}
Choose a finite partition of \([0,1]\) of mesh at most
\(\delta\), and round each \(\eta\) to its lower and upper
endpoints \(\eta_-\) and \(\eta_+\). They obey
\(\eta_-\leq\eta\leq\eta_+\) and
\(\eta_+-\eta_-\leq\delta\).
The rounded-down channel is obtained from the actual one by
local attenuation at Bob, using the disclosed record. Conversely,
the rounded-up model simulates the actual channel: draw the
residual \(\eta\) from its conditional law within the bin,
share it publicly at the prescribed time, and attenuate by
\(\eta/\eta_+\). The all-zero bin causes no difficulty.
The residual draw is public randomness independent of the input
before it is disclosed.

For all \(x\geq0\), the entropy function satisfies
\begin{equation}
0\leq c_{\eta_+}(x)-c_{\eta_-}(x)
\leq g(\delta x).
\label{fd:eq:quantization}
\end{equation}
This follows from \(g(u+v)-g(u)\leq g(v)\).
Thus the receiver formulas of the two rounded channels differ
by at most \(g(\delta N)\).

For advance information, optimizing over allocations that vary
within a bin is equivalent to optimizing over their conditional
means when its endpoint transmissivity is fixed: concavity of
\(c_\eta\) can only improve the objective upon this averaging.
For any admissible allocation, the gap of the endpoint objectives
is bounded by
\begin{equation}
\int g(\delta x(\eta))\,\mu(\dd\eta)
\leq g(\delta N).
\end{equation}
The optimized gap has the same bound. The finite-alphabet
operational bounds consequently squeeze both continuous-record
capacities to their stated integrals as \(\delta\downarrow0\).
This argument also accommodates atoms, including at zero and
one, and avoids assigning an entropy to a continuous classical
record.

\section{Optimal energy allocation for advance fading information}
\label{fd:app:allocation}
For \(x>0\), direct differentiation gives
\begin{equation}
c_\eta''(x)=
-\frac{\eta}{\ln2\,x(1+x)[1+(1-\eta)x]}.
\label{fd:eq:secondderivative}
\end{equation}
For \(\eta>0\), \(c_\eta\) is therefore strictly concave,
\(c_\eta'(0+)=+\infty\), and \(c_\eta'(\infty)=0\).
Every \(\lambda>0\) defines a unique \(x_\lambda(\eta)>0\)
through Eq.~\eqref{fd:eq:allocation}. This solution is measurable,
decreases continuously with \(\lambda\), and satisfies
\begin{equation}
0\leq x_\lambda(\eta)\leq\frac1{2^\lambda-1},
\end{equation}
because \(c_\eta'(x)\leq g'(x)\).
Dominated convergence on compact positive ranges of \(\lambda\)
proves continuity of \(\int x_\lambda\,\dd\mu\).
This integral tends to zero as \(\lambda\to\infty\) and
to infinity as \(\lambda\downarrow0\), provided
\(\mu((0,1])>0\). There is consequently a unique multiplier
with mean allocation \(N>0\).

The supporting-line inequality gives, for any allocation \(y\),
\begin{equation}
c_\eta(y)-c_\eta(x_\lambda)
\leq\lambda(y-x_\lambda),~~ \eta>0.
\end{equation}
It also holds at \(\eta=0\) with \(x_\lambda(0)=0\).
Integrating proves global optimality under the mean budget.
Strict concavity proves uniqueness on the positive-transmissivity
set. No energy is optimally assigned to \(\eta=0\), and the
entire budget is used if a positive-transmissivity set has
nonzero probability.

Finally, for \(0<\eta<1\) and \(z=(1-\eta)x\),
\begin{equation}
\partial_\eta c_\eta'(x)
=\log_2(1+z^{-1})-\frac1{\ln2(1+z)}>0.
\end{equation}
Together with Eq.~\eqref{fd:eq:secondderivative}, this shows that
\(x_\lambda(\eta)\) increases with \(\eta\). The endpoints
follow by continuity. Infinite slope at zero explains why this
mean-only allocation populates every positive transmissivity.

\section{The loss-count extension and its adaptive converse}
\label{fd:app:extension}
\subsection{Construction and entropy functional}
Write the pure-loss Kraus operators as
\begin{equation}
L_{\eta,\ell}\ket n
=\begin{cases}
\sqrt{\binom n\ell}\eta^{(n-\ell)/2}(1-\eta)^{\ell/2}\ket{n-\ell},
&n\geq\ell,\\
0,&n<\ell.
\end{cases}
\label{fd:eq:kraus}
\end{equation}
The extension
\begin{equation}
\Fhat(\rho)=\sum_{\ell\geq0}\int
L_{\eta,\ell}\rho L_{\eta,\ell}^{\dagger}\,\mu(\dd\eta)
\otimes\proj\ell_F
\label{fd:eq:lossflag}
\end{equation}
reveals the photon-loss count to Bob, but does not reveal \(\eta\).
Tracing out \(F\) recovers \(\F\).

For \(n\geq\ell\), recall the vectors in \(L^2(\mu)\)
introduced in Sec.~\ref{fd:sec:comparison}:
\begin{equation}
v_{n\ell}(\eta)=
\sqrt{\binom n\ell}\eta^{(n-\ell)/2}(1-\eta)^{\ell/2}.
\label{fd:eq:vectors}
\end{equation}
Use \(v_{n\ell}=0\) for \(\ell>n\), and the usual continuous
interpretation at the endpoints. Let
\begin{align}
b_{n\ell}&=\|v_{n\ell}\|^2,~~
\tau_n=\bigoplus_{\ell\geq0}\proj{v_{n\ell}},\label{fd:eq:taun}\\
\Gamma_\ell(\pi)_{nm}
&=\sqrt{\pi_n\pi_m}\langle v_{m\ell}|v_{n\ell}\rangle,
~~ n,m\geq\ell.\label{fd:eq:gram}
\end{align}
The \(\tau_n\) are normalized states, and \(b_n=(b_{n\ell})_\ell\)
is the loss-count distribution for an \(n\)-photon input. For a
photon distribution \(\pi\), put
\begin{align}
\Bfun(\pi)&=\sum_n\pi_n H(b_n),\label{fd:eq:B}\\
\Afun(\pi)&=H(\pi)+\Bfun(\pi)
-S\!\left(\bigoplus_\ell\Gamma_\ell(\pi)\right).\label{fd:eq:A}
\end{align}
Equivalently,
\begin{equation}
\Afun(\pi)=H(X|E)_{\Omega^\pi},~~
\Omega^\pi_{XE}=\sum_n\pi_n\proj n_X\otimes\tau_n.
\label{fd:eq:cqentropy}
\end{equation}
This representation proves concavity and nonnegativity of \(\Afun\).
All entropies are finite at finite input energy; in particular
\(0\leq\Afun(\pi)\leq H(\pi)\leq g(N)\).

The comparison channel has the exact assisted capacities
\begin{equation}
\begin{aligned}
Q_2(\Fhat,N)&=D_2(\Fhat,N)\\
&=P_2(\Fhat,N)=K_2(\Fhat,N)\\
&=U_\mu(N).
\end{aligned}
\label{fd:eq:extensioncapacity}
\end{equation}
We prove this identity and the bounds of Theorem~\ref{fd:thm:extension}
below.

\subsection{Dilation and finite entropies}
The Hilbert space
\(\mathcal H_E=\bigoplus_{\ell\geq0}L^2(\mu)\) is separable.
An isometry for \(\Fhat\) is
\begin{equation}
V\ket n=\sum_{\ell=0}^n
\ket{n-\ell}_B\ket\ell_F\ket{v_{n\ell}}_E,
\label{fd:eq:flagdilation}
\end{equation}
where the final vector lies in the \(\ell\)th summand.
Vectors with different \(n\) have orthogonal \(BF\) labels,
and \(\sum_\ell\|v_{n\ell}\|^2=1\), so \(V\) is indeed
an isometry. Tracing \(E\) gives Eq.~\eqref{fd:eq:lossflag};
tracing \(BF\) gives
\begin{equation}
\Fhat^c(\rho)=\sum_n\rho_{nn}\tau_n.
\label{fd:eq:flagcomplement}
\end{equation}
This is a classical-to-quantum, entanglement-breaking complement
\cite{HolevoEB2008}.

For fixed \(n\), the state \(\tau_n\) has one pure vector in
each orthogonal loss-count summand, hence
\(S(\tau_n)=H(b_n)\). Realize the loss count as the sum of
\(n\) binary variables which are conditionally independent given
\(\eta\). Their unconditional marginal loss probability is
\(1-\bar\eta\), and therefore
\begin{equation}
H(b_n)\leq nh_2(\bar\eta).
\label{fd:eq:Bbound}
\end{equation}
At input mean at most \(N\), it follows that
\begin{align}
H(\pi)&\leq g(N),~~ \Bfun(\pi)\leq Nh_2(\bar\eta),\nonumber\\
S\!\left(\sum_n\pi_n\tau_n\right)
&\leq H(\pi)+\Bfun(\pi)<\infty.
\label{fd:eq:entropycontrol}
\end{align}
The nonzero spectrum of \(\sum_n\pi_n\tau_n\) is that of
\(\bigoplus_\ell\Gamma_\ell(\pi)\), by the Gram-matrix
identity. This proves Eq.~\eqref{fd:eq:cqentropy} and justifies
every entropy difference used in defining \(\Afun\).

\subsection{Number pinching with an arbitrary reference}
Let \(\Delta_B\) pinch the output in the photon-number basis,
acting trivially on \(F\) and all other systems. Let $\Delta$
be complete photon-number dephasing at the input and put
\(\mathcal M=\Fhat\circ\Delta\). Then
\begin{equation}
\mathcal M=\Delta_B\circ\Fhat.
\label{fd:eq:commutingpinching}
\end{equation}
For a pure state with arbitrary reference \(R\), write
\begin{equation}
\ket\psi_{RA}=\sum_n\sqrt{\pi_n}\ket{r_n}_R\ket n_A,
\end{equation}
where the normalized \(\ket{r_n}\) need not be orthogonal.
Within a fixed \(\ell\) block of the output, however,
the vectors \(\ket{r_n}\ket{n-\ell}\) are orthonormal.
The output block consequently has matrix \(\Gamma_\ell\).
Its pinched state has entropy \(H(\pi)+\Bfun(\pi)\).
The relative-entropy identity for pinching gives
\begin{equation}
D\bigl((\id_R\otimes\Fhat)(\psi)\|
(\id_R\otimes\mathcal M)(\psi)\bigr)=\Afun(\pi).
\label{fd:eq:purecoherence}
\end{equation}
Purify a mixed \(\rho_{RA}\) and trace out the extra reference.
Data processing then gives the inequality
\begin{equation}
D\bigl((\id_R\otimes\Fhat)(\rho)\|
(\id_R\otimes\mathcal M)(\rho)\bigr)\leq\Afun(\pi_\rho).
\label{fd:eq:mixedcoherence}
\end{equation}
No entropy bound on the original reference is needed. In the
purified calculation, both entropies whose difference appears
in Eq.~\eqref{fd:eq:purecoherence} are finite by
Eq.~\eqref{fd:eq:entropycontrol}.

For states \(\rho,\sigma\) including the reference, the
pinching identity and data processing imply
\begin{align}
D(\Fhat(\rho)\|\mathcal M(\sigma))
={}&D(\Fhat(\rho)\|\mathcal M(\rho))\nonumber\\
&+D(\mathcal M(\rho)\|\mathcal M(\sigma))\nonumber\\
\leq{}&\Afun(\pi_\rho)+D(\rho\|\sigma).
\label{fd:eq:fullchain}
\end{align}
Identity maps on references are suppressed. The pinching identity
is valid in the extended sense; for countably many photon
projectors it follows from finite partitions and monotone
limits of relative entropy. The first term is finite by
Eq.~\eqref{fd:eq:mixedcoherence}; an infinite second term makes
the inequality immediate. This handles arbitrary separable
reference spaces without subtracting two infinite entropies.

\subsection{Entanglement increase and exact extension capacity}
Take \(\rho_{XAY}\), with \(XA\) at Alice and \(Y\) at
Bob, and a separable comparison state \(\sigma_{XA:Y}\).
The map \(\mathcal M\) measures input photon number and
prepares a diagonal state of \(BF\). Thus
\((\id_{XY}\otimes\mathcal M)(\sigma)\) is separable
across \(X:BFY\). Infimizing Eq.~\eqref{fd:eq:fullchain}
over \(\sigma\) gives
\begin{equation}
\ER(X:BFY)_{\Fhat(\rho)}
\leq\ER(XA:Y)_\rho+\Afun(\pi_\rho).
\label{fd:eq:flagreegain}
\end{equation}
This is a one-round bound on unregularized REE. Iterating it
through a complete adaptive protocol, using LOCC monotonicity,
gives \(\ER(\zeta_n)\leq\sum_i\Afun(\pi_i)\).
Conditional entropy is concave, and the mean distribution
\(\bar\pi=n^{-1}\sum_i\pi_i\) has energy at most \(N\).
Hence
\begin{equation}
\ER(\zeta_n)\leq n\Afun(\bar\pi)\leq nU_\mu(N).
\label{fd:eq:flagadaptive}
\end{equation}
The target tests prove the quantum and key converses.

For achievability, purify a Fock-diagonal input distribution
\(\pi\). Its \(BF\) output is classical with joint law
\(\pi_nb_{n\ell}\), whose entropy is
\(H(\pi)+\Bfun(\pi)\). The joint reference-output entropy
is \(S(\bigoplus_\ell\Gamma_\ell)\). The coherent
information is exactly \(\Afun(\pi)\), achievable by
one-way hashing of the distributed states.

Finite-support distributions suffice for this achievability.
To see this, let \(\epsilon=\sum_{n>M}\pi_n\), and let
\(\pi_{\leq M}\) and \(\pi_{>M}\) be the normalized
restrictions. Replace the tail by vacuum:
\(\widetilde\pi=(1-\epsilon)\pi_{\leq M}+\epsilon\delta_0\).
This does not increase mean energy. The conditional-entropy
chain rule with the classical tail indicator gives
\begin{align}
\Afun(\pi)
\leq{}&h_2(\epsilon)+(1-\epsilon)\Afun(\pi_{\leq M})
+\epsilon H(\pi_{>M}),\nonumber\\
\Afun(\widetilde\pi)
\geq{}&(1-\epsilon)\Afun(\pi_{\leq M}).
\end{align}
Since \(H(\pi_{>M})\leq g(N/\epsilon)\) and
\(\epsilon\leq N/(M+1)\), the difference is at most
\(h_2(\epsilon)+\epsilon g(N/\epsilon)\to0\).
Finite-support inputs have finite-dimensional effective outputs,
so ordinary hashing applies before this limit is taken.
This proves Eq.~\eqref{fd:eq:extensioncapacity}.

\subsection{Returning to the unobserved channel}
For the purification
\(\ket{\psi_\pi}=\sum_n\sqrt{\pi_n}\ket n_R\ket n_A\),
the output of \(\F\) is a direct sum over loss counts:
the \(\ell\)th block is supported by
\(\{\ket n_R\ket{n-\ell}_B:n\geq\ell\}\) and has
matrix \(\Gamma_\ell(\pi)\). Orthogonality of these joint
subspaces does not supply Bob with a local loss-count record.
Nevertheless it determines the entropy, giving
\begin{equation}
I_{\rm R}(\pi,\F)
=H(\pi)-S\!\left(\bigoplus_\ell\Gamma_\ell(\pi)\right)
=\Afun(\pi)-\Bfun(\pi).
\label{fd:eq:rciidentity}
\end{equation}
Use finite-support distributions approaching \(U_\mu(N)\)
and Eq.~\eqref{fd:eq:Bbound} to obtain
\(Q_2(\F,N)\geq U_\mu(N)-Nh_2(\bar\eta)\), with zero
as an additional trivial lower bound. Tracing the loss flag
and disclosing the full fading record, respectively, give
\(K_2(\F,N)\leq U_\mu(N)\) and
\(K_2(\F,N)\leq\Crec(N)\).
These prove Eq.~\eqref{fd:eq:bracket}. If \(U_\mu(N)\) is
smaller than \(Nh_2(\bar\eta)\), the upper bound alone
implies Eq.~\eqref{fd:eq:keygap}; otherwise subtract the stated
lower bound. The conclusion therefore holds in both cases.

\section{Single-photon rail bound for fading}
\label{fd:app:rail}

A complementary achievable rate uses one photon over \(m\) rails.
Set \(r=\beta/\bar\eta\) when \(\bar\eta>0\), and define
\begin{align}
z_m&=r+(1-r)/m,\nonumber\\
d_m(r)&=\log_2m-h_2(z_m)-(1-z_m)\log_2(m-1).
\end{align}
Then
\begin{equation}
Q_2(\F,N)\geq
L_\mu(N):=\sup_{m\geq2}
\bar\eta\min\{N,m^{-1}\}\,d_m(r).
\label{fd:eq:rail}
\end{equation}
For $\bar\eta=0$ the bound is zero. We now derive the rate.

Over \(m\) channel uses, prepare
\(m^{-1/2}\sum_{j=1}^m\ket j_R\ket{1_j}\), where
\(\ket{1_j}\) has one photon in rail \(j\). Bob measures
only the total output photon number. The acceptance probability
is \(\bar\eta\). A diagonal rail term survives with factor
\(\bar\eta\), whereas coherence between distinct rails
survives with factor \(a^2=\beta\), because their fading
variables are independent. The accepted state is therefore
\begin{equation}
\rho_{\rm acc}=r\proj{\Phi_m}
+\frac{1-r}{m}\sum_{j=1}^m\proj{jj},
~~ r=\beta/\bar\eta.
\end{equation}
Its nonzero eigenvalues are \(z_m=r+(1-r)/m\) and
\((1-r)/m\), the latter with multiplicity \(m-1\).
Hashing gives \(d_m(r)\) ebits per accepted block, hence
\(\bar\eta d_m(r)/m\) per channel use. The energy is
\(1/m\) per use. Time sharing with vacuum when \(N<1/m\)
gives Eq.~\eqref{fd:eq:rail}. This recovers the known
single-photon fading rail expression
\cite[Appendix B]{Catalano2026}, now with explicit accounting
for time sharing under the unconditional budget.

\section{On/off fading: scalar reductions and numerics}
\label{fd:app:binary}
\subsection{Entropy identities}
For \(\Phi_p=p\,\id+(1-p)\mathcal L_0\), write an input
distribution as
\begin{equation}
\pi_0=1-\epsilon,~~ \pi_n=\epsilon w_n\quad(n\geq1).
\end{equation}
For its number-correlated purification the channel output is
\begin{equation}
\rho_{RB}=p\proj{\psi_\pi}
+(1-p)\sum_n\pi_n\proj{n0}.
\label{fd:eq:onoffstate}
\end{equation}
For each positive \(n\), \(\ket{n0}\) is orthogonal to
\(\ket{\psi_\pi}\) and to \(\ket{00}\).
The remaining two-dimensional block is spanned by
\(\ket{00}\) and \(\sum_{n\geq1}\sqrt{w_n}\ket{nn}\).
It has trace \(t_\epsilon\), determinant
\(p(1-p)\epsilon(1-\epsilon)\), and eigenvalues
\(t_\epsilon-s_\epsilon,s_\epsilon\).
The other eigenvalues are \((1-p)\epsilon w_n\).
Thus
\begin{align}
S(\rho_{RB})={}&H(t_\epsilon-s_\epsilon,s_\epsilon,(1-p)\epsilon)
\nonumber\\
&+(1-p)\epsilon H(w),\nonumber\\
I_{\rm R}(\pi,\Phi_p)={}&j_p(\epsilon)+p\epsilon H(w).
\label{fd:eq:onoffentropy}
\end{align}
For \(n\geq1\), the loss count is zero with probability
\(p\) and \(n\) with probability \(1-p\). Hence
\(\Bfun(\pi)=\epsilon h_2(p)\) and
\begin{equation}
\Afun(\pi)=j_p(\epsilon)+\epsilon h_2(p)+p\epsilon H(w).
\label{fd:eq:onoffA}
\end{equation}

For fixed \(\epsilon>0\), the conditional mean is at most
\(M=N/\epsilon\geq1\). Both objectives maximize \(H(w)\),
whose optimum is \(g(M-1)\), attained by
\begin{equation}
w_n=\frac1M\left(1-\frac1M\right)^{n-1},~~ n\geq1.
\label{fd:eq:shiftedgeometric}
\end{equation}
At \(M=1\) this is \(\delta_1\). Substitution proves
Eqs.~\eqref{fd:eq:onoffU} and \eqref{fd:eq:onoffR} for diagonal
inputs. The endpoint \(\epsilon\downarrow0\) has value
zero, since \(\epsilon g(N/\epsilon-1)\to0\).
The maxima therefore exist on the displayed compact interval.

\subsection{Why diagonal inputs suffice for reverse hashing}
For any channel \(\mathcal N\), reverse coherent information
as a function of the input is
\(I_{\rm R}(\rho,\mathcal N)=S(\rho)-S(\mathcal N^c(\rho))\).
For an input ensemble its concavity gap is
\begin{equation}
\chi(\{p_j,\rho_j\})-
\chi(\{p_j,\mathcal N^c(\rho_j)\})\geq0,
\end{equation}
by data processing. Thus \(I_{\rm R}\) is concave
\cite{Reverse}. The on/off channel is phase covariant,
and reverse coherent information is invariant under input
phase rotations. Phase averaging preserves energy and can
only increase \(I_{\rm R}\), producing a Fock-diagonal
input. The required entropies are finite at finite energy:
the output mean is \(pN\) and a complementary entropy is
at most \(g(N)+g(pN)\). Finite-support approximation or
finite phase averages therefore justify the same argument
in the bosonic setting. This completes the identification
of \(R_p(N)\) as the optimized single-use reverse coherent
information, without claiming that it is the two-way capacity.

\subsection{An additional erasure-code lower bound}
On the strictly positive-photon subspace, a blocked transmission
produces the orthogonal vacuum flag. The restriction of
\(\Phi_p\) is consequently an erasure channel of success
probability \(p\) \cite{BennettErasure1997}. A diagonal state of mean \(M\geq1\)
in this subspace can have entropy \(g(M-1)\), and
two-way entanglement distribution attains \(p g(M-1)\).
Use this code in a fraction \(t\) of all slots and vacuum
in the rest. Taking \(M=N/t\) gives the lower bound
\begin{equation}
E_p(N)=\max_{0\leq t\leq\min\{N,1\}}
pt\,g(N/t-1),
\label{fd:eq:nonvacuum}
\end{equation}
with value zero at \(t=0\). A useful closed form is
\begin{equation}
E_p(N)=\begin{cases}
pN,&0\leq N\leq2,\\
p\,g(N-1),&N\geq2.
\end{cases}
\label{fd:eq:nonvacuumclosed}
\end{equation}
Indeed, \(t g(N/t-1)=Nh_2(t/N)\), whose unconstrained
maximum occurs at \(t=N/2\); for \(N>2\) the constraint
\(t\leq1\) fixes the optimum at \(t=1\).

\subsection{Reproducible numerical evaluation}
The supplied script evaluates Eqs.~\eqref{fd:eq:onoffU} and
\eqref{fd:eq:onoffR} by scalar bounded optimization, with endpoint
checks. It compares these with the closed observed capacities,
Eq.~\eqref{fd:eq:nonvacuum}, and a finite set of rail dimensions.
A maximum over the evaluated rail dimensions is a valid lower
bound without any assertion that the infinite rail optimization
has been exhausted. Figure~\ref{fd:fig:capacities} uses 200
logarithmically spaced energies from \(10^{-4}\) to \(30\)
at \(p=1/2\). The data, plotting code, and vector figure
are provided with the source. Scalar optimizations use standard
floating-point arithmetic; the mathematical bounds themselves
are the variational expressions proved above.

\section{Strong-converse thresholds for fading}\label{fd:app:strong}

\subsection{Conditioning on the fading sequence}

We first prove the upper threshold when $\Bfade<\infty$. Then
$\eta<1$ almost surely. Let
$\boldsymbol\eta=(\eta_1,\ldots,\eta_n)$ denote the full sequence
of independent draws and put
\begin{equation}
 S_n(\boldsymbol\eta)=\sum_{i=1}^{n}b(\eta_i).
 \label{fd:eq:randomsum}
\end{equation}
Conditioning on this sequence gives a finite ordered list of pure-loss
channels. The bound may allow the parties to know the entire sequence
before transmission; this only increases their resources. All inputs
and local operations remain arbitrary and adaptive. Since the
unconditional mean transmitted energy is finite, the conditional input
means at the finitely many slots are finite for almost every sequence.
They need not obey any sequencewise uniform bound.

The finite-product testing argument of Appendix~\ref{app:meanstrong} applies
to this list with one simulation resource per call. In particular, for conditional
squared fidelity $f\in(0,1)$ to a quantum or private-state target
of dimension $d_n$,
\begin{equation}
 \log_2d_n\leq S_n(\boldsymbol\eta)
       +\log_2 6+2\log_2\frac{2-f}{f}.
 \label{fd:eq:conditionalstrong}
\end{equation}
Indeed, the relative entropies of the component resources add, and
their limiting variances vanish. Both simulation and regularization
are removed with this finite protocol and its sequence fixed, as in
Appendices~\ref{app:meanstrong} and~\ref{app:continuousstrong}. The endpoint $\eta_i=0$
is an entanglement-breaking call with zero contribution. Equation
\eqref{fd:eq:conditionalstrong} also holds at $f=1$ by taking the
error parameter to zero. Rearranging and using $2-f\leq2$ gives
\begin{equation}
 f\leq\min\left\{1,\sqrt{24}\,
 2^{-[\log_2d_n-S_n(\boldsymbol\eta)]/2}\right\}.
 \label{fd:eq:conditionalfidelity}
\end{equation}
This includes $f=0$ and requires no photon bound uniform in the
sequence, blocklength, or fading support.

Write $R_n=n^{-1}\log_2d_n$ and fix $\delta>0$ such that
$R_n\geq\Bfade+2\delta$ for all sufficiently large $n$. The atypical
probability
\begin{equation}
 q_n(\delta)=\Prb\{S_n>n(\Bfade+\delta)\}
       \longrightarrow0
 \label{fd:eq:typicalfading}
\end{equation}
vanishes by the law of large numbers, since the nonnegative random
variable $b(\eta)$ has finite first moment. For quantum targets,
squared fidelity to the fixed maximally entangled state is linear in
the output state. Averaging \eqref{fd:eq:conditionalfidelity} therefore
yields
\begin{equation}
 f_n\leq q_n(\delta)+\sqrt{24}\,2^{-n\delta/2}
       \longrightarrow0.
 \label{fd:eq:averagedfidelity}
\end{equation}
This proof requires neither a finite variance for $b(\eta)$ nor a
uniform separation of the fading support from $\eta=1$.

For operational key, the public fading sequence is part of Eve's
record. Let $e_n(\boldsymbol\eta)$ be the conditional trace-distance
error, with the actual conditional Eve marginal retained in the ideal
comparison. Classical disintegration gives
$e_n=\int e_n(\boldsymbol\eta)\,\mu^{\otimes n}(\dd\boldsymbol\eta)$.
The energy-preserving key-to-private-state reduction gives conditional
private-state fidelity at least $[1-e_n(\boldsymbol\eta)]^2$.
Equation \eqref{fd:eq:conditionalfidelity} then implies
\begin{equation}
 1-e_n\leq q_n(\delta)+24^{1/4}\,2^{-n\delta/4}
       \longrightarrow0.
 \label{fd:eq:averagedkeyerror}
\end{equation}
Finite quantizations of a Borel record give the same conditional
formulation without assigning an entropy to the record.
Teleportation and the one-time pad transfer the conclusions to the
other two tasks under the conventions of \cref{sec:model}.

The upper bound for unobserved fading follows by the receiver-record
extension of the same channel. Discarding the legitimate receiver's
flag recovers the averaged optical map; the environment of the
extension is obtainable by a channel on an environment of the
unobserved map. Thus an unobserved protocol is also admissible in the
recorded model, with the same quantum output and with key error no
larger after the environment is degraded. Equations
\eqref{fd:eq:averagedfidelity} and \eqref{fd:eq:averagedkeyerror}
therefore give \eqref{fd:eq:generalstrongupper}.

\subsection{High-energy limits and a matching threshold}

For observed fading, monotone convergence applied to the exact
receiver formula gives
\begin{equation}
 \lim_{M\to\infty}\Crec(M)
 =\int\lim_{M\to\infty}c_\eta(M)\,\mu(\dd\eta)
 =\Bfade,
 \label{fd:eq:observedhighenergy}
\end{equation}
including an infinite limit. Since a constant allocation is allowed
with advance information and $c_\eta(x)\leq b(\eta)$,
$\Crec(M)\leq\Cpre(M)\leq\Bfade$; the advance capacity has the
same limit.

For unobserved fading with finite support, the result of
Ref.~\cite[Theorem 1 and Appendix B]{Catalano2026} gives
\begin{equation}
 \lim_{M\to\infty}Q_2(\F,M)
 =\lim_{M\to\infty}K_2(\F,M)=\Bfade.
 \label{fd:eq:hiddenhighenergy}
\end{equation}
The stated unconstrained capacities in that reference are explicitly
these limits of mean-energy-constrained capacities. Thus, for every
fixed $R<\Bfade$, its coding construction supplies a vanishing-error
code at some finite mean energy $M$. All displacement energy needed
to identify the component is included in $M$. The finite support is
essential to this cited lower bound; we do not extend it to a
general unobserved Borel distribution.

For either observed model, or for unobserved finite-support fading,
fix $N>0$ and $0<R<\Bfade$. Choose a finite $M>N$ such that an
active code has rate tending to $R$ and error tending to zero. With
public probability $\theta=N/M$, run this code; otherwise send
vacuum in all the same slots and output a fixed product state or a
fixed public key. Both branches have the same target dimension $d_n$.
The unconditional energy is at most $N$, while
\begin{equation}
 f_n=\theta f_n^{\mathrm{on}}+(1-\theta)/d_n
       \longrightarrow\theta>0
 \label{fd:eq:fadingdilution}
\end{equation}
for a quantum target. For key, the orthogonal public flag gives
$e_n=\theta e_n^{\mathrm{on}}+(1-\theta)(1-1/d_n)
\to1-\theta<1$. Hence no threshold below $\Bfade$ is possible.
Combined with the upper bound, this proves
\cref{fd:eq:observedstrong,fd:eq:hiddenstrong} when the average is
finite. If $\Bfade=+\infty$, the same construction works at every
finite $R$ and proves an infinite threshold without requiring an
upper-bound argument. At $N=0$, every transmitted optical mode is
vacuum and the initial and public resources cannot generate a
positive-rate quantum or secret-key target.

For the on/off channel with $p>0$, the divergence needed for the
lower threshold also follows from the explicit erasure-code rate
$E_p(M)=p\,g(M-1)$ for $M\geq2$ in
\eqref{fd:eq:nonvacuumclosed}. This proves its infinite threshold
directly. Finally, when the channel is nonzero, the attained observed
finite-energy capacities are strictly below $\Bfade$ because
$c_\eta(x)<b(\eta)$ for every finite $x$ and every $\eta>0$.
The unobserved finite-support capacity is no larger than
$\Crec(N)$. These facts and the dilution argument explain the
failure of a strong converse at the finite-energy capacities in the
cases covered by the exact threshold identities.

\end{document}